\documentclass{aa520}
\usepackage[english]{babel}
\usepackage{latexsym}
\usepackage{epsfig}
\usepackage{psfig}
\usepackage{amssymb}
\usepackage{amsfonts}
\usepackage{graphicx}
\usepackage{natbib}

\newcommand{\enum}{\begin{enumerate}}
\newcommand{\eenum}{\end{enumerate}}
\newcommand{\bgeq}{\begin{equation}}
\newcommand{\eneq}{\end{equation}} \newcommand{\de}{\mbox{d}}

\newcommand{\mdot}{\dot{M}} \newcommand{\uno}{({\it i}) }
\newcommand{\due}{({\it ii}) } \newcommand{\tre}{({\it iii}) }
 \newcommand{\msun}{M_{\odot}}
\newcommand{\msunyr}{M_{\odot}/\mbox{yr}}
\newcommand{\micron}{\mu\mbox{m}}

\defcitealias{kenyon88}{KHH}
\defcitealias{popham96}{PKHN}
\defcitealias{LB2001}{LB}
\defcitealias{BL99}{BL}
\bibpunct{(}{)}{;}{a}{,}{,}

\begin{document}

\title{Probing the rotation curve of the outer accretion disk in
FU Orionis objects with long-wavelength spectroscopy}

\titlerunning{Probing the rotation curve of FU Orionis objects}

\author{G. Lodato \inst{1} \and G. Bertin \inst{2}}

\institute{Insitute of Astronomy, University of Cambridge, Madingley
Road, CB3 0HA, Cambridge, UK \and Universit\`a degli Studi di Milano,
Dipartimento di Fisica, via Celoria 16, I-20133 Milano, Italy}

\date{Received date/Accepted date} 

\abstract{Studies of the Spectral Energy Distribution of Young Stellar
Objects suggest that the outer disk of FU Orionis objects might be
self-gravitating. In this paper we propose a method to test directly
whether, in these objects, significant deviations from Keplerian
rotation occur. In a first approach, we have used a simplified model
of the disk vertical structure that allows us to quickly bring out
effects related to the disk self-gravity. We find that the often
studied optical and near-infrared line profiles are produced too close
to the central object to provide significant evidence for
non-Keplerian rotation. Based on parameters relevant for the case of
FU Ori, we show that high-resolution long-wavelength spectroscopy, of
the far-infrared H$_2$ pure rotational lines (sometimes observed in
``passive'' protostellar disks) and sub-mm CO lines, should be well
suited to probe the rotation curve in the outer disk, thus measuring
to what extent it is affected by the disk self-gravity. The results of
the present exploratory paper should be extended soon to a more
realistic treatment of the disk vertical structure.

}

\maketitle

\section{Introduction}

Very young (Class 0 or Class I) protostellar sources are thought to be
characterized by a fairly high mass accretion rate \citep{calvet2000},
but they appear to be still deeply embedded in their protostellar
envelopes, so that their disk remains hidden to direct
observations. On the other hand, in the more evolved T Tauri stars the
mass accretion rate is generally considered to be modest, so that
their disks are heated by the combined contribution of viscous
dissipation and irradiation from the central star, making it more
difficult to extract detailed information about the accretion
process. In this context, FU Orionis objects \citep{hartmann96} are a
rather small but remarkable class of pre-main sequence stars, because
they are ideal ``laboratories'' to test the process of disk accretion
during the early stages of star formation. In fact, the disk of FU
Orionis objects, differently from that of Class I objects, can be
studied directly from its optical emission, and, differently from that
of T Tauri stars, is likely to be the site of ``active" accretion.

The distinctive feature of FU Orionis objects is that they undergo a
violent outburst phase. During the outburst, they can increase their
luminosity by more than 4 magnitudes in a period of a few years, and
then slowly return to a quiescent state on a much longer timescale. It
is commonly believed that such outbursts are the result of a
significant increase of the mass accretion rate in the disk which
usually surrounds Young Stellar Objects. Many possible mechanisms to
trigger the outburst phase have been discussed in the literature,
among which are a tidal interaction with a companion star
\citep{bonnell92}, the onset of a thermal instability in a partially
ionized disk, when the outer disk is already accreting at a
sufficiently high rate \citep{belletal95}, or the onset of thermal
instability induced by the presence of a satellite within the disk
\citep{clarkesyer96}. The high accretion rate makes the emission of
these systems dominated by the accretion luminosity, with only a minor
contribution from the central star.

The modeling of the Spectral Energy Distribution (SED) of FU Orionis
objects provides an estimate of the product $M_{\star}\mdot$ (as
discussed below in Sect. \ref{sec:tempprof}) which turns out to be of
the order of $10^{-4}\msun^2$/yr. For a stellar mass
$M_{\star}\approx\msun$, this would correspond to an accretion rate as
high as $\approx 10^{-4}\msunyr$. ``Standard'' models based on the
presence of a Keplerian accretion disk lead to a satisfactory fit of
the available photometric data only for wavelengths shorter than
$10\micron$; the luminosity at longer wavelengths is usually much
larger than expected. A flaring disk model, in which the outer disk is
illuminated by the inner disk, could in principle explain the
far-infrared excess, but the required amount of flaring turns out to
be often too large \citep{kenyon91}; therefore, the long-wavelength
part of the SED is generally attributed to an infalling envelope,
which is heated by the accretion disk luminosity.

Recently we have shown the viability of the picture (\citealt{LB2001};
hereafter LB) in which the long wavelength SED of FU Orionis objects
is considered to be the signature of the effects of the disk
self-gravity. There are already many clues that point to the
importance of the disk self-gravity in this context. Among these, we
may recall that: \uno Submillimeter observations \citep{sandell2001}
show that the disk masses in these systems are much higher than those
of T Tauri stars (indicating that they are probably younger than their
low-luminosity counterparts); \due The modeling of the outburst in
terms of a thermal instability event \citep{bellin94} requires a very
low value for the viscosity parameter $\alpha$, which in turn leads to
a high surface density; indeed, detailed vertical structure
calculations \citep{belletal97} show that at low values of the
viscosity the disk is marginally stable against axisymmetric
gravitational disturbances already at a distance significantly smaller
than 1 AU from the central star. In fact, it has been suggested that
FU Orionis outbursts could be triggered by a mechanism that is based
on the role of the disk self-gravity \citep{armitage2001}.

In the model proposed by \citetalias{LB2001} the far infrared excess
is produced by two separate contributions: \uno The higher rotation
curve of the disk, in the case where the disk is sufficiently massive,
enhances the viscous dissipation rate; \due The extra heating required
to keep the outer disk marginally stable with respect to gravitational
instabilities makes the surface temperature higher at large radii.

The second contribution is probably related also to non-local energy
transport processes that should become important when global
gravitational instabilities are present. The importance of global
effects in self-gravitating disks has been brought out by means of
numerical simulations \citep{rice03}, but further work is desired to
clarify the issue of non-local transport in massive disks (Lodato \&
Rice, in preparation). Thus, for such second contribution, much of the
discussion focuses on theoretical aspects \citep{balbus99}.

The importance of the first contribution can be checked by means of
direct observations. In this paper we address the issue of detecting
non-Keplerian rotation in protostellar disks. In passing, we note that
the non-Keplerian rotation observed in some AGN accretion disks can be
successfully explained by the model on which this paper is based
\citep{LB03}. Here we propose a possible measurement of the rotation
of the outer disk in FU Orionis objects based on the analysis of
mid-infrared and sub-millimetric spectroscopy. While sub-millimetric
emission from FU Orionis objects is generally considered to come from
the disk \citep{weintraub91}, alternative scenarios are available for
the interpretation of the far-infrared emission. In fact, as we have
already mentioned, \citet{kenyon91} attribute the far infrared
emission to the envelope. Here, following \citetalias{LB2001}, we will
assume that the far infrared emission comes from the disk.

A useful diagnostics to probe the kinematical properties of FU Orionis
disks is provided by the shape of the observed line profiles. One
merit of this kind of diagnostics resides in the fact that the surface
temperature of the disk decreases with radius, and that the wavelength
characterizing the emission from a certain annulus of the disk changes
accordingly. Therefore, the study of the profiles of different lines
allows us to probe the disk kinematics at different radial distances
from the central accreting protostar.  In particular, the observed
optical and near-infrared line profiles are usually double-peaked
\citep[ hereafter KHH]{kenyon88}, as expected from a rotating disk; in
addition, the peak separation decreases with increasing wavelength, as
expected for a disk with a decreasing temperature profile \citep[
hereafter PKHN]{popham96}.  We can thus anticipate that from the peak
separation at very long wavelengths in the far infrared and in the
sub-mm, probing the outer disk, we should be able to tell whether the
rotation is significantly different from Keplerian, hence testing the
predictions of a model in which the disk self-gravity affects the
shape of the rotation curve.

This kind of observations will require high spectral resolution (in
order to detect small velocity differences) and high spatial
resolution (in order to disentangle different kinematical
comnponents). In fact, many protostellar disks are known to possess
strong molecular outflows. Therefore, a possible source of confusion
in the shape of the line profiles may come from the kinematics of the
outflows, that could result in asymmetries in the shape of the line
profile.  In this respect, the high spatial resolution that can be
achieved with interferometric techniques \citep{guilloteau98}, and, in
the near future, with ALMA will eventually lead to clear-cut tests on
the role of disk self-gravity in the rotation curve of protostellar
disks.

The paper is organized as follows: in Section \ref{procedure} we
describe our method to test the rotation properties of the outer
accretion disk; in Section \ref{fu} we describe the basic properties
of the adopted accretion disk model based on parameters relevant for
FU Ori; in Section \ref{optical} we describe our expectations for
optical and NIR line profiles; in Section \ref{mir} we consider the
case of mid-infrared pure rotational H$_2$ line profiles; in Section
\ref{radio} we describe the expected shape of sub-mm CO line emission;
in Section \ref{conclusion} we draw our conclusions.

\section{Measuring deviations from Keplerian rotation in spatially
unresolved protostellar disks}
\label{procedure}

The method presented in this Section is an extension of the method
described by \citetalias{kenyon88}. Our main assumption is that the
SED of FU Orionis objects is produced by an optically thick accretion
disk up to wavelengths of the order of $100\micron$, consistent with
studies of the millimetric continuum in these systems
\citep{weintraub91}. The basic procedure can then be summarized as
follows:

\begin{enumerate}

\item The surface temperature profile $T_s(r)$ is obtained from a
parametric fit to the SED; note that in this paper we are not
interested in explaining the physical origin of this profile, which we
addressed in a previous article (LB). Given a value of the inclination
angle $i$, this fit leads to a determination of the inner and outer
radii of the disk, $r_{in}$ and $r_{out}$, and of the product $\mdot
M_{\star}$, where $\mdot$ is the mass accretion rate and $M_{\star}$
is the mass of the central object.

\item The profiles of the optical-NIR lines (such as some CO
absorption bands, seen in FU Orionis objects), which are produced
mostly in the inner disk, are used to determine the rotational
velocity in the inner Keplerian disk. For a given value of the
inclination $i$, fitting the observed line shapes yields a measurement
of $M_{\star}/r_{in}$. As a result of the first two steps, we obtain
separate estimates of $M_{\star}$ and $\mdot$. So far, the procedure
is very close to that of \citetalias{kenyon88}.

\item In significantly massive disks, the line profiles at longer
wavelengths should be broader than expected from a Keplerian
extrapolation of the rotation curve measured in the previous
step. Fitting the observed mid-infrared and sub-millimeter line
profiles with the self-gravitating disk model described in
\citet[ hereafter BL]{BL99} will then provide a value for the
lengthscale $r_s$ that marks the transition to the non-Keplerian
rotation regime (see Subsection \ref{sec:rotation} below). This fit
will test directly the importance of the disk self-gravity,
independently of the arguments that can be made in order to explain
the infrared excess in the SED. Furthermore, since $r_s$ depends on
$M_{\star}$, $\mdot$, and on the viscosity parameter $\alpha$ (see
Eq. (\ref{rs}) below), this fit will also lead to a measurement of the
viscosity parameter $\alpha$.

\end{enumerate}

\subsection{Determining the radial temperature profile}
\label{sec:tempprof}

We start by introducing the dimensionless radial coordinate
$x=r/r_{in}$ and the temperature scale $T_0$ from the Keplerian model
temperature profile:
\begin{equation}
\label{tprof1}
\sigma_BT_s^4(r)=\frac{3}{8\pi}\frac{GM_{\star}\mdot}{r_{in}^3}
\left(\frac{r_{in}}{r} \right)^3=\frac{\sigma_BT_0^4}{x^3},
\end{equation}
where $\sigma_B$ is the Stefan-Boltzmann constant.  This simple
temperature profile is able to reproduce the correct spectral index of
the SED at wavelengths smaller than 10$\micron$ and thus is adequate
at small radii.

In the following, we would like to test the self-gravitating disk
hypothesis by studying directly the rotation properties of the disk,
independently of a discussion of the detailed physical processes that
justify the observed SED. Therefore we consider a parametric
description of the temperature profile, in line with many other
investigations of YSO disks (e.g., see \citealt{becksarg93} and
\citealt{osterloh95}, who studied the sub-mm continuum of YSOs, or
\citealt{thi2001}, who studied emission lines). For the purpose, we
thus consider a simple combination of two power laws, one describing
the inner disk (and with temperature therefore proportional to
$r^{-3/4}$), and the other describing the outer disk, allowing for
deviations from the $r^{-3/4}$ dependence:
\begin{equation}
\label{tprof2}
T_s^4(r)=T_0^4\theta^4(x),
\end{equation}
where
\begin{equation}
\theta^4(x)=\frac{1}{x^{3}}\left[1+\left(\frac{x}{x_t}\right)^{\beta}\right],
\end{equation}
$x_t=r_t/r_{in}$ being a dimensionless radius, beyond which the
temperature profile departs from the ``standard'' profile, and $\beta$
an appropriate power-law index.

Under the assumption that the disk emission is optically thick, the
SED can be constructed by integrating in radius the blackbody
spectrum. By analogy with what described in \citetalias{LB2001}, we
then introduce a reference frequency $\nu_0=kT_0/h$ and rescale the
frequency to $\hat{\nu}=\nu/\nu_0$. For a given disk inclination $i$,
the SED is thus given by:
\begin{equation}
4\pi D_0^2\nu F_{\nu}=L_0\hat{\nu}^4\int_{1}^{x_{out}}\frac{x\de x}
{\mbox{e}^{\hat{\nu}/\theta(x)}-1},
\end{equation}
where
\begin{equation}
\label{l0}
L_0=\frac{120}{\pi^3}r_{in}^2\sigma_BT_0^4\cos i,
\end{equation}
$D_0$ is the distance to the YSO, and $x_{out}=r_{out}/r_{in}$.

A fit to the SED of observed objects will therefore determine the fit
parameters $x_{out}$, $T_0$, $L_0$, $\beta$, and $x_t$. The combined
knowledge of $L_0$ and $T_0$ (assuming the inclination angle of the
disk to be known) allows us to obtain $r_{in}$ and the product
$M_{\star}\mdot$ (see Eqs. (\ref{tprof1}) and (\ref{l0})).

\subsection{The rotation curve in the presence of a self-gravitating disk}
\label{sec:rotation}

We now consider the contribution of the disk self-gravity to the
rotation curve. Based on the results of \citetalias{BL99}, we
introduce a typical lengthscale $r_s$, the radius beyond which
deviations from Keplerian rotation become significant, and a typical
velocity scale $V_{0}$, defined as follows:
\begin{equation}
\label{rs}
r_s=2GM_{\star}\left(\frac{\bar{Q}}{4}\right)^2\left(\frac{G\mdot}
{2\alpha}\right)^{-2/3},
\end{equation}
\begin{equation}
V_{0}^2=\frac{8}{\bar{Q}^2}\left(\frac{G\mdot}{2\alpha}\right)^{2/3}.
\end{equation}
Here $\bar{Q}$ is a parameter that, for simplicity, will be set to
unity (see \citetalias{BL99}). The quantity $\alpha$ is the parameter
that defines the \citet{shakura73} viscosity prescription. The
rotation curve of self-regulated, self-gravitating accretion disks is
given by:
\begin{equation}
\label{vsg}
V^2=\frac{GM_{\star}}{r}+V_{0}^2+V_{0}^2\hat{\phi}^2(r/r_s)=
V_{\star}^2\left[\frac{1}{x}+\frac{1}{x_s}(1+\hat{\phi}^2(x/x_s))\right],
\end{equation}
where $V_{\star}^2=GM_{\star}/r_{in}$, $x_s=r_s/r_{in}$ and
$\hat{\phi}^2$ is defined in Eq. (4) of \citetalias{BL99}.

In the following, we will sometimes use the short, simplified
expression ``self-gravitating rotation curve" (as opposed to Keplerian
rotation curve) when we will refer to the rotation curve derived from
Eq. (\ref{vsg}) for a model that includes the effects of the disk
self-gravity.

\subsection{The shape of the line profiles}
\label{sec:line}

In the present analysis we focus on the line shape. Therefore, we will
consider only the relative contribution of the different annuli of the
disk to the line intensity.

We will express the line profiles as a function of the velocity shift
$\Delta v$, corresponding to the frequency shift $\Delta\nu$ from the
line center $\nu_l$, so that $\Delta v/c=\Delta\nu/\nu_l$. The Doppler
line profile $\phi(\Delta v,r)$ at a given radius of a disk rotating
with velocity $V(r)$ is given by:
\begin{displaymath}
\label{lineprofile}
\phi(\Delta v,r)=\left\{
\begin{array}{ll}
\displaystyle\frac{c}{\sqrt{V_{los}^2(r)-(\Delta v)^2}} &
\textrm{if $(\Delta v)^2<V_{los}^2(r)$}\\
0                                    & \textrm{otherwise}

\end{array}
\right.
\end{displaymath}
where $V_{los}(r)=V(r)\sin i$ is the maximum line-of-sight rotational
velocity. Obviously, this expression takes into account only the
effects related to the disk ordered motions; intrinsic effects
associated with the microscopic physics of line emission and with the
thermal motions inside the disk are not included.

We convolve this line profile with a Gaussian profile
$\mbox{e}^{-(\Delta v/v_b)^2}$, either to simulate the effects of a
finite instrumental spectral resolution $R$, so that $v_b/c=1/R$, or
to describe turbulent or thermal motions of the emitting species.  The
global line profile will be given by integrating the convolved line
profile $\tilde{\phi}(\Delta v, r)$ over radius, weighted by an
appropriate weight function $l(r,\nu_l)$ which expresses the relative
contribution of the different annuli of the disk to the line emission:
\begin{equation}
\label{globallineprof} F(\Delta v)=\int_{r_{in}}^{r_{out}}
l(r,\nu_l)\tilde{\phi}(\Delta v, r) 2\pi r\de r~.
\end{equation}

The function $l(r,\nu_l)$ is specified once the details of the
emission or of the absorption are known. At wavelengths where the
continuum emission is optically thin, we expect to find emission
lines; on the other hand, where the continuum is optically thick,
emission lines are produced only if a temperature inversion in the
upper layers of the disk is present, due, for example, to the effect
of irradiation from the central object \citep{calvet91}. If we
consider FU Orionis objects, since the accretion rate is high and
irradiation is not expected to be efficient, we thus expect to have
absorption lines at optical and infrared wavelengths
\citep{calvet93}. Near infrared CO absorption bands from the disk are
indeed a typical feature in FU Orionis objects
\citepalias{kenyon88}. However, irradiation from the inner disk may
give rise to a hot atmosphere above the disk, where emission lines can
be produced. Therefore, in some cases (for example, for the optically
thin FIR H$_2$ molecular lines, see Section \ref{mir}) we have
considered both cases of absorption and emission.

In this simplified analysis we will consider for emission lines the
two limiting cases of optically thick (which can be appropriate for
sub-millimetric CO rotational lines) and optically thin emission (best
suited for FIR pure rotational H$_2$ emission lines, see below), so as
to keep the radiative transport as simple as possible.
\begin{enumerate}
\item {\bf Optically thick emission lines.} In this case the
brightness temperature of the emission line coincides with the
physical temperature $T_s(r)$ (see \citealt{becksarg93}) so that we
can assume that the different annuli of the disk emit proportionally
to their contribution to blackbody emission at the wavelength of the
line center:
\begin{equation}
\label{thick}
l(r,\nu_l)\propto\frac{1}{\mbox{e}^{h\nu_l/kT_s(r)}-1}.
\end{equation}
The expected line profile can thus be obtained directly, based on the
results of modeling of the SED, as described in
Sec. \ref{sec:tempprof}.

\item {\bf Optically thin emission lines.} In this case the relative
contribution of the different annuli requires one additional piece of
information, about the surface density radial profile $\sigma(r)$.
Apart from numerical factors that do not depend on the radial
coordinate or on the temperature, we can express $l(r,\nu_l)$ as:
\begin{equation}
\label{thin}
l(r,\nu_l)\propto\sigma(r)\frac{\mbox{e}^{-E_{up}/kT_s(r)}}{Q(T_s(r))},
\end{equation}
where $E_{up}$ is the energy of the upper level of the transition, and
$Q(T_s(r))$ is the partition function of the line emitting species
(for example the H$_2$ molecule; see also \citealt{thi2001}). For
simplicity, in this formula we have made the assumption that the line
emitting species is at the same temperature as the matter responsible
for the continuum emission (i.e. mostly the dust in the upper layers
of the disk). However, it should be noted that this assumption would
lead to very weak emission lines, difficult to observe. This has been
partially confirmed by FIR observations of pure rotational H$_2$
emission lines in quiescent YSOs \citep{richter2002b}, that indeed
reveal that these emission lines, if present at all, have a very weak
line strength.

A more detailed discussion of the radiative transfer would require a
complete accretion disk model. In the present simplified
investigation, in order to proceed further from Eq.~(\ref{thin}), we
may consider a surface density profile of the form $\sigma \sim
r^{-\gamma}$. The power law index $\gamma$ can be derived from
spatially resolved observations of disks at mm wavelengths, where the
disk emission is optically thin \citep{guilloteau98}, and it turns out
to be often close to unity \citep{testi03,pietu03} (even for evolved
objects, such as TW Hya; see \citealt{wilner2000}). Therefore, the
case $\gamma=1$ will be our reference case (see also the discussion is
Section \ref{robustness} below). [Interestingly, the self-gravitating
disk solution of \citetalias{BL99} indeed predicts $\gamma=1$, but, as
we did when we introduced the properties of the SED, in this paper we
do not wish to address the problem of interpreting the observed
profile.]

\item{\bf Absorption lines.} In this case we follow
\citetalias{popham96} and make the simplifying assumption that the
equivalent width of the line does not change with $T_s$, so that
$l(r,\nu_l)$ is proportional to the blackbody emission and will be
therefore described by Eq. (\ref{thick}), as for optically thick
emission lines. As \citetalias{popham96} note, this approach may not
be valid above some threshold temperature (dependent on the absorption
line), where the relevant molecules producing absorption disappear. We
have therefore set the equivalent width to zero above this
temperature (see also Sec. \ref{optical} and \ref{mir} below). 

Other models of the absorption lines in FU Orionis include the use of
stellar template photospheres with different temperatures, simulating
the disk radial temperature profile (the disk is divided into rings,
and with each ring a stellar template spectrum is associated,
characterized by effective temperature equal to the disk surface
temperature at that radius; \citetalias{kenyon88}), or of the detailed
solution of the radiative transfer in the disk atmosphere
\citep{calvet93}.  However, such refined calculations show only minor
differences with respect to estimates based on the simplest
assumptions, at least for optical and near-infrared lines. More
caution should be taken when applying the simple constant equivalent
width assumption to longer wavelength lines (such as the H$_2$ lines
that we are going to consider in Sec \ref{mir}). In this first
analysis we have decided to follow the simplified approach, being
aware that for the final comparison with observations more realistic
assumptions about the radiative transfer in the disk should be
considered.

\end{enumerate}

\subsection{Other factors expected to affect the shape of the line profiles}

Before concluding this Section, a cautionary remark is in order.  The
main objective of the present paper is to show quantitatively what
kind of spectroscopic observations could carry significant evidence
for the presence of non-Keplerian motions in protostellar disks.  In
view of this goal, we have decided to consider a simplified
description of the vertical structure and of the radiative transfer in
the disk. Obviously, for the subsequent goal of interpreting the
observations of some specific objects, many other effects concerning
the dynamics, the physics and the chemistry of protostellar disks
should be considered carefully. These include the possible interplay
between rotation and infall in determining the shape of the line
profile \citep{hoger01,belloche02}, the effect of molecular depletion
(see \citealt{markwick02} and comment at the end of Section
\ref{radio}), and the effect of other sources of gravitational field
in disturbing the rotation curve. This final issue can be particularly
important in the case of FU Orionis objects, for which the tidal
interaction with a close ``companion'' has been sometimes suggested
\citep{bonnell92} as a possible way to trigger the outburst. However,
event statistics \citep{hartmann96} suggest that FU Orionis might be a
repetitive phenomenon, with a recurrence period of $\approx 10^4-10^5$
yrs. The ``companion'' should therefore lie in a very eccentric orbit,
so that it should cause only a minor disturbance to the shape of the
line profiles after the outburst has been triggered.

\section{Basic parameters for a model suggested by the SED of FU Ori}
\label{fu}

In this Section we apply the initial step of the procedure outlined
above to the SED of FU Ori, the prototypical FU Orionis
object. \citetalias{LB2001} have already modeled the SED of this
object in terms of a self-gravitating accretion disk, obtaining a
satisfactory fit to the available data up to $100\micron$. Therefore,
this object is a good candidate for testing its rotation
characteristics and possible deviations from Keplerian rotation.  In
Table \ref{tab:fuor} we report the basic physical parameters for FU
Ori obtained from the modeling of its SED in previous analyses
(i.e. \citetalias{kenyon88}, \citetalias{popham96}, and
\citetalias{LB2001}) and from the parametric modeling described in
Subsect.~2.1 and thus adopted in this work (see below).

\begin{table*}[hbt!]
\centering

\caption{Accretion disk parameters for FU Ori, for the disk models of
KHH, PKHN, LB, and this work. Values marked with an asterisk are
assumed.}

\begin{tabular}{cccccccc}
\hline
\hline
 & $M_{\star}/M_{\odot}$ & $\mdot/(\msunyr)$ & $M_{\star}\mdot/
(M_{\odot}^2/yr)$ & $r_{in}/R_{\odot}$ & $\cos i$ &  $A_V$ & $D_0$/pc\\
\hline
KHH  & $0.34$  & $4.36~10^{-4}$ &$1.4~10^{-4}$ & 5.4 & $0.5^*$
& $1.85$  & $500^*$\\
PKHN & $0.7$   & $2~10^{-4}$    &$1.4~10^{-4}$ & 5.7 & $0.5$
& $2.2^*$ & $500^*$\\
LB   & $1^*$   & $0.82~10^{-4}$ &$0.82~10^{-4}$& 8   & $0.65$
& $2^*$   & $550^*$\\
this work & $0.7^*$ & $1.4~10^{-4}$    &$1~10^{-4}$ & 11  & $0.5^*$
& $2.2^*$   & $500^*$\\

\hline

\end{tabular}

\label{tab:fuor}
\end{table*}

The different models cannot be easily compared to each other, because
of the different parameters assumed (distance, extinction and
inclination angle) and of the different modeling considered for the
emission in the disk boundary layer. In particular, the main
characteristics of the models are:
\begin{itemize}
\item[-] KHH assume that the net angular momentum flux in the disk is
not vanishing and is given by the requirement of zero torque at the
inner disk radius. The boundary layer is assumed to be at a constant
temperature, corresponding to the maximum disk temperature.
\item[-] PKHN assume that the net angular momentum flux in the disk is
vanishing and include the boundary layer emission, treating it as a
low-radiative-efficiency flow.
\item[-] LB (and this work) assume that the net angular momentum flux
in the disk is vanishing, but do not include the boundary layer
emission.
\end{itemize}

The differences in the choice of the visual extinction $A_V$ (in
magnitudes) and of the distance $D_0$ only induce slight changes in
the estimate of the product $M_{\star}\mdot$, when similar values for
the inclination are assumed. On the other hand, the different
treatment of the inner disk emission leads to a different estimate of
the inner disk radius in the three models, so that
$r_{in}^{\mathrm{KHH}}\lesssim r_{in}^{\mathrm{PKHN}}<r_{in}^
{\mathrm{LB}}$. We interpret the relatively large value of the inner
radius estimated by LB and in this work as a consequence of the fact
that, with respect to PKHN, we did not include the contribution of the
radiatively inefficient boundary layer, so that our inner disk radius
is to be considered closer to the transition from the thin disk
solution to the hot boundary layer, rather than to the actual stellar
radius. Another important difference between KHH and PKHN concerns the
central mass estimate: the boundary layer in the PKHN model rotates
slower than Keplerian, due to the effect of the increased pressure
gradient in the boundary layer. Thus, to obtain the same rotational
velocity in the inner disk (to match the optical line profiles; see
below) a larger central mass is required.

We will follow the work of PKHN as much as possible. Therefore, we
start by assuming the extinction coefficient, distance and inclination
angle provided in their analysis. In particular, the inclination angle
assumed below is such that $\cos{i}=0.5$, slightly smaller than the
value $\cos{i}=0.65$ derived by LB.

\begin{figure}
  \resizebox{\hsize}{!}{\includegraphics{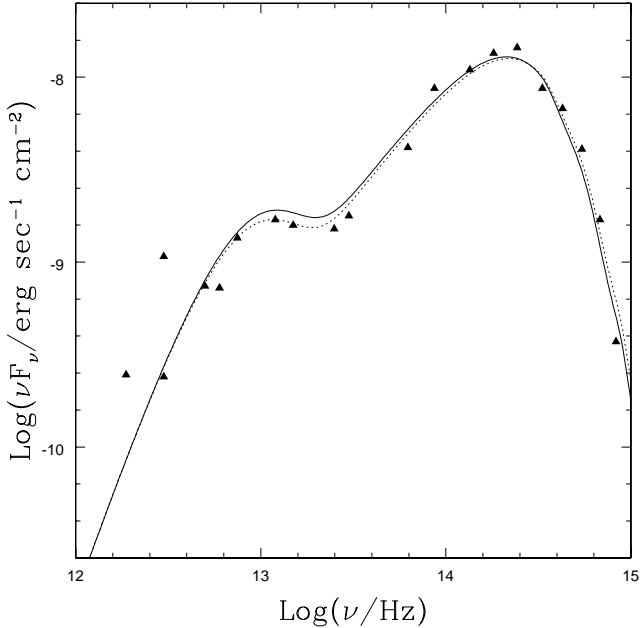}}
\caption{Spectral Energy Distribution of FU Ori. The triangles show
  the data (from \citealt{kenyon91}), while the curves show the best
  fit model obtained in this work (solid line) and by LB (dotted
  line).}
\label{fig:fuor}
\end{figure}

\begin{figure}
  \resizebox{\hsize}{!}{\includegraphics{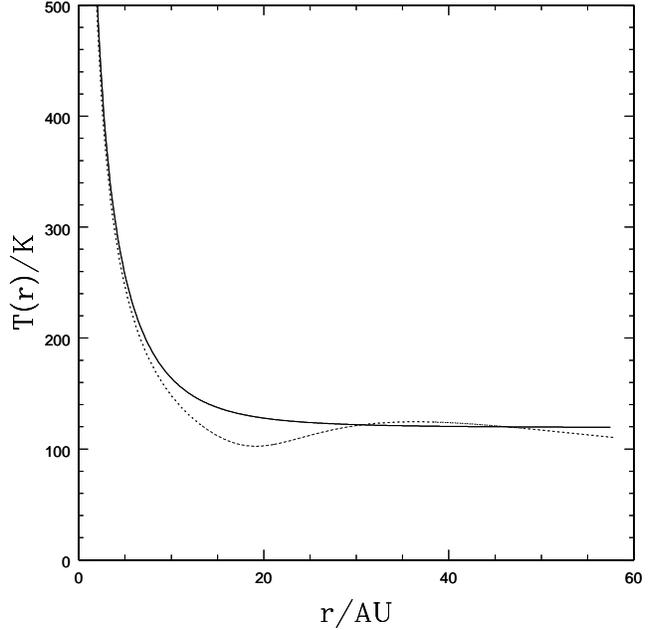}}
\caption{Surface temperature profile of FU Ori obtained in this
work from a parametric model of the SED (solid line), compared to
that obtained by LB (dotted line).} \label{fig:tprof}
\end{figure}

With the parametric model described in Section \ref{sec:tempprof} we
have fitted the available photometric data (from \citealt{kenyon91}),
corrected by means of a standard dereddening function
\citep{cardelli89}. The derived values for the product
$M_{\star}\mdot$ and $r_{in}$ are shown in Table \ref{tab:fuor}.
Figure \ref{fig:fuor} shows the observed SED of FU Ori together with
our best fit (solid line) and the best fit obtained by LB with a
self-gravitating disk model (dotted line). The best fit values for
$x_t$ and $\beta$ (see Section \ref{sec:tempprof}) are: $x_t\approx
261$ (which corresponds to $r_t\approx 13$ AU) and $\beta=3$.  In view
of the goals of this paper, of the inhomogeneity of the SED data, and
especially of the detailed discussion provided in KHH, PKHN, and LB,
we will not comment here on the issue of setting uncertainties in the
value of the derived parameters recorded in Table 1. We only note that
the relative poor quality of the SED data might result is some
ambiguities in the values of the estimated parameters, as already
noted by \citet{thamm94}.

Note that $\beta=3$ implies that the surface temperature profile is
flat in the outer disk. In Fig.  \ref{fig:tprof} the temperature
profile of our best fit parametric model (solid line) is displayed
together with the best fit model profile obtained by LB (dotted line).

\section{Optical-NIR line profiles}
\label{optical}

\begin{figure}
  \resizebox{\hsize}{!}{\includegraphics{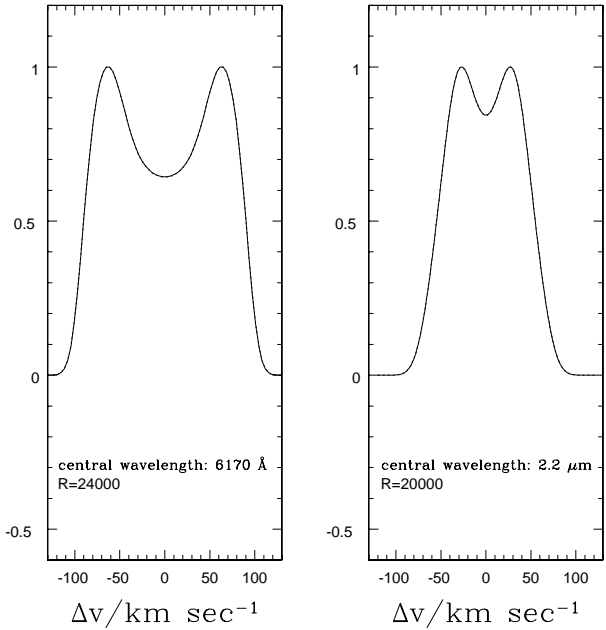}}
\caption{Optical (6170\AA) and near-IR (2.2$\micron$) predicted
line profiles for FU Orionis, in the Keplerian (solid line) and
self-gravitating (dotted line) model, with $r_s\approx 18$ AU.
The curves practically overlap and no differences between the two
descriptions are noted.} \label{fig:nir}
\end{figure}

The contribution of the disk mass to the rotation curve depends on the
size of the transition radius $r_s$ (see Eq.~(\ref{rs})) in relation
to the region of the disk actually probed by the available
spectroscopy. In this Section we will show that, in the case of a
model suggested by the study of the SED of FU Ori, $r_s$ is large
enough so that the disk self-gravity has little or no influence on the
optical-NIR line profiles, which are mostly produced in the inner
disk. In practice, from the peak separation of the observed line
profiles at these relatively short wavelengths, one can measure the
ratio $M_{\star}/r_{in}$, thus preparing the ground for a firm
expectation of effects related to deviations from Keplerian rotation
in spectroscopic studies at larger wavelengths (see following
Section). This Section thus addresses step 2 of the procedure outlined
in Sec. \ref{procedure}.

Here we follow PKHN and assume their value for the central mass,
$M_{\star}=0.7\msun$ (PKHN found significantly larger values of
$M_{\star}$, with respect to KHH who did not include a detailed model
of the inner boundary layer). Assuming $\alpha=0.034$ (the best fit
value obtained by LB), we would obtain $r_s\approx 350r_{in}\approx
18$ AU. Actually, as a result of the different estimates of
$M_{\star}$ and $\mdot$ considered here with respect to LB, this value
for $r_s$ is smaller than that obtained by LB, leading to a higher and
possibly unrealistic total disk mass ($\approx 2 M_{\star}$). We have
thus also considered the case in which $\alpha=0.1$, leading to
$r_s\approx 700 r_{in}\approx 36$ AU, better in line with the best
model of LB.

We refer to the same (absorption) lines as in PKHN and KHH, i.e. metal
lines at 6170 \AA (at a resolution $R=24000$) and CO lines at
$2.2\micron$ (at a resolution $R=20000$; the spectral resolutions are
taken to be the same as in PKHN).  The threshold temperatures above
which the equivalent width of the line is taken to be vanishing are
8000K for the optical lines, and 5000K for the near-infrared lines. In
Fig. \ref{fig:nir} we plot, for $r_s\approx 18$ AU, both the profile
resulting from strictly Keplerian rotation (solid line) and the one
resulting from the self-gravitating model (dotted line). The two
different profiles are clearly indistinguishable. Note that the
quantity usually referred to is not the actual absorption line
profile, but the peak of the cross-correlation function of many
absorption lines in the same wavelength range. This quantity is
preferred from an observational point of view because it gives a way
to obtain higher signal-to-noise ratios; physically, it represents the
average shape of the line profile in a small wavelength range.

A similar plot for the case in which $r_s \approx 36$ AU would again
show, as expected, no significant differences in the shape of the line
profiles between the strictly Keplerian model and the self-gravitating
disk model.

\section{Probing the outer disk rotation with H$_2$ line profiles}
\label{mir}

In order to probe properties of the outer disk, so as to have a chance
for detecting significant deviations from Keplerian rotation, one
should consider line profiles at longer wavelengths (see
Fig. \ref{fig:tprof}).

Molecular hydrogen pure rotational emission lines have been observed
with the Infrared Space Observatory (ISO) in many protostellar systems
\citep{vandichoek98,thi2001}. However, due to the low angular
resolution of ISO, it is not clear whether the H$_2$ emission can be
actually associated with the disk. Indeed, these ISO observations were
not confirmed at higher spatial resolution \citep{richter2002b}. The
weakness of H$_2$ emission lines could indicate that spatial and
temperature separations between the gas and the dust in the disk are
small. Molecular hydrogen is a homonuclear molecule, so that the
rotational emission lines result from the electric
quadrupole. Therefore, they have a very low Einstein $A$ coefficient,
which makes them optically thin up to high column densities (of the
order of $10^{23}\mbox{cm}^{-2}$). We will therefore assume that these
lines are optically thin, although we should be aware that, if the
self-gravitating disk hypothesis is correct, then the surface density
of the disk is going to be high as well, so that the optically thin
assumption should eventually be re-examined. The basic line parameters
of the H$_2$ rotational lines are summarized in Table \ref{tab:h2}.

\begin{table*}[hbt!]
\centering
\caption{Pure rotational H$_2$ line parameters: central wavelength,
energy of the upper level (in Kelvin), Einstein $A$-coefficient. From
\citet{thi2001}.}

\begin{tabular}{cccc}
\hline
\hline
 transition & Wavelength ($\micron$) & $E_{up}$ (K) & $A$-coefficient
(sec$^{-1}$)\\
\hline

S(0) $2\rightarrow 0$  &  28.218     &  509.88      &$2.94~10^{-11}$\\
S(1) $3\rightarrow 1$  &  17.035     &  1015.12     &$4.76~10^{-10}$\\
S(2) $4\rightarrow 2$  &  12.278     &  1814.43     &$2.76~10^{-9}$ \\
S(3) $5\rightarrow 3$  &  9.662      &  2503.82     &$9.84~10^{-9}$ \\

\hline

\end{tabular}

\label{tab:h2}
\end{table*}

\begin{figure*}
\centerline{\epsfig{figure=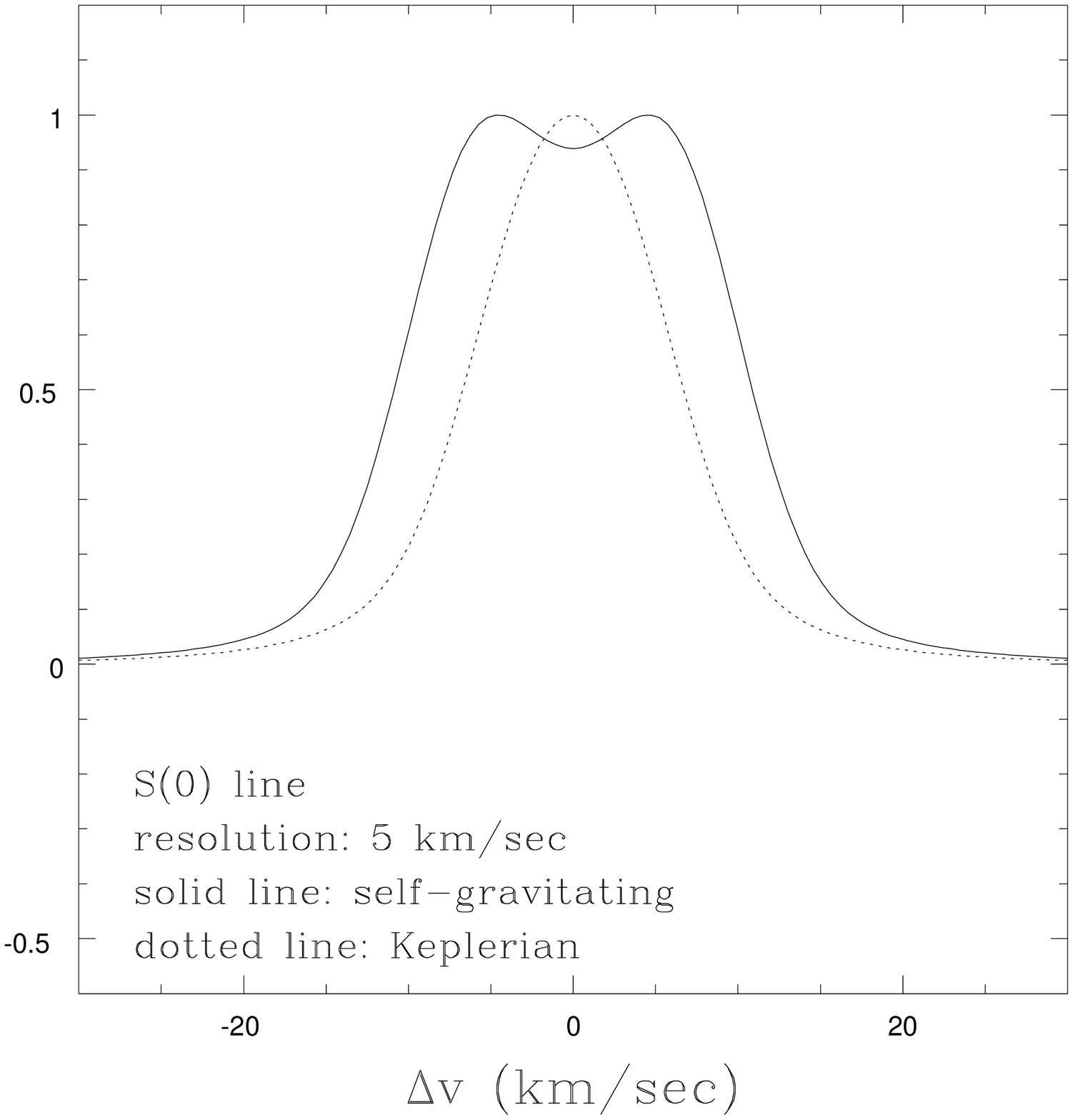,height=7.5cm}
            \epsfig{figure=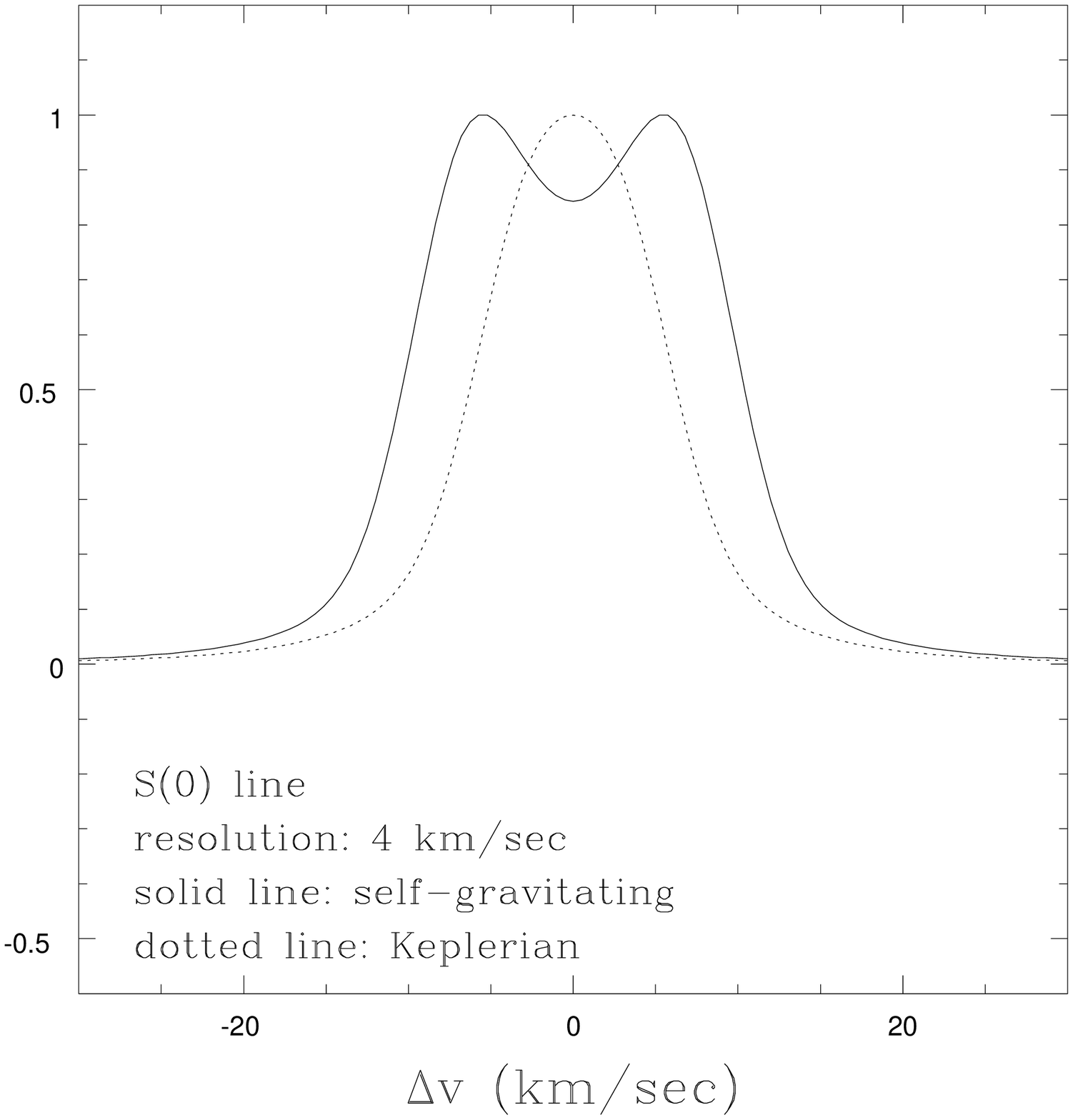,height=7.5cm}}
\centerline{\epsfig{figure=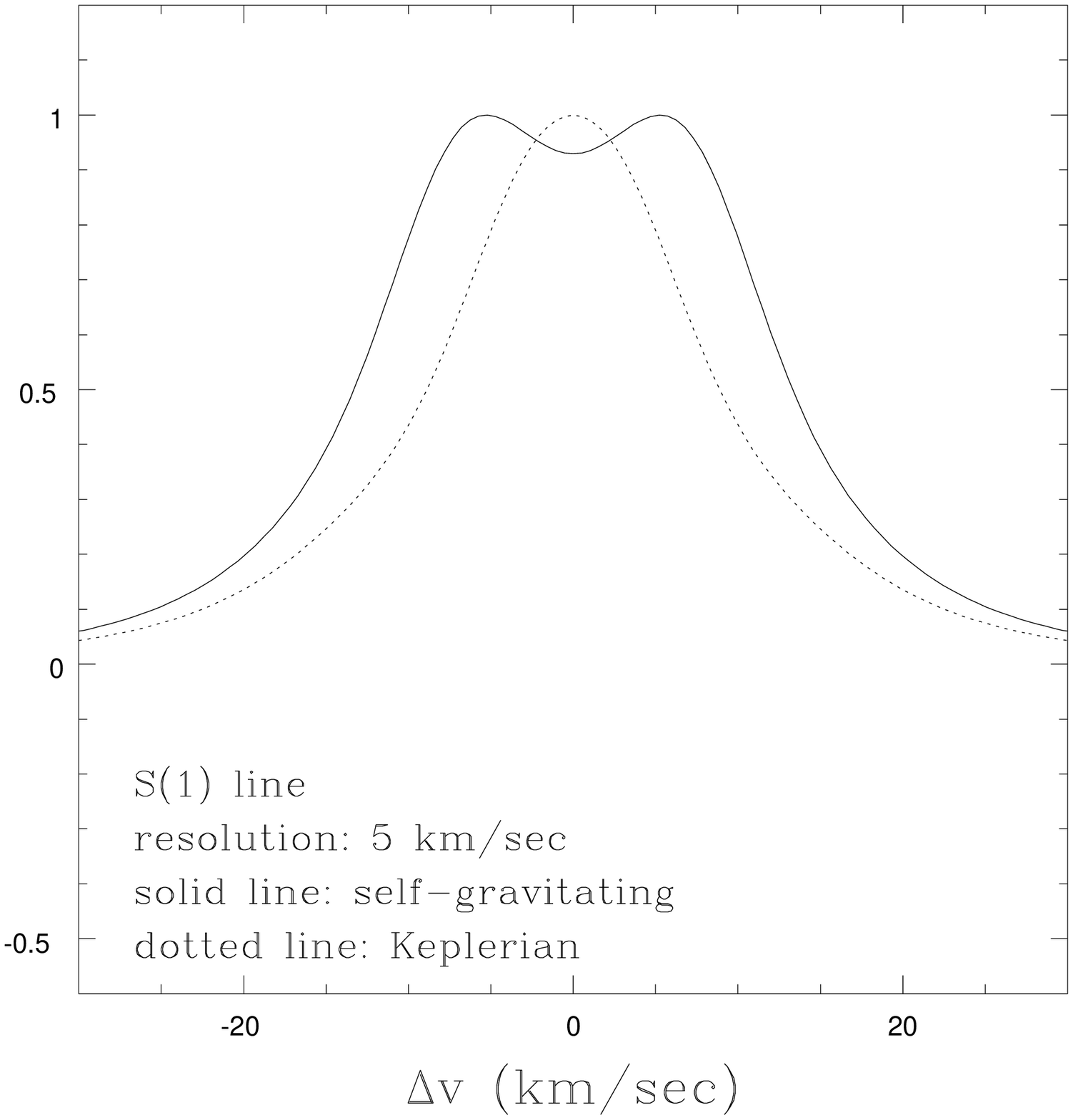,height=7.5cm}
            \epsfig{figure=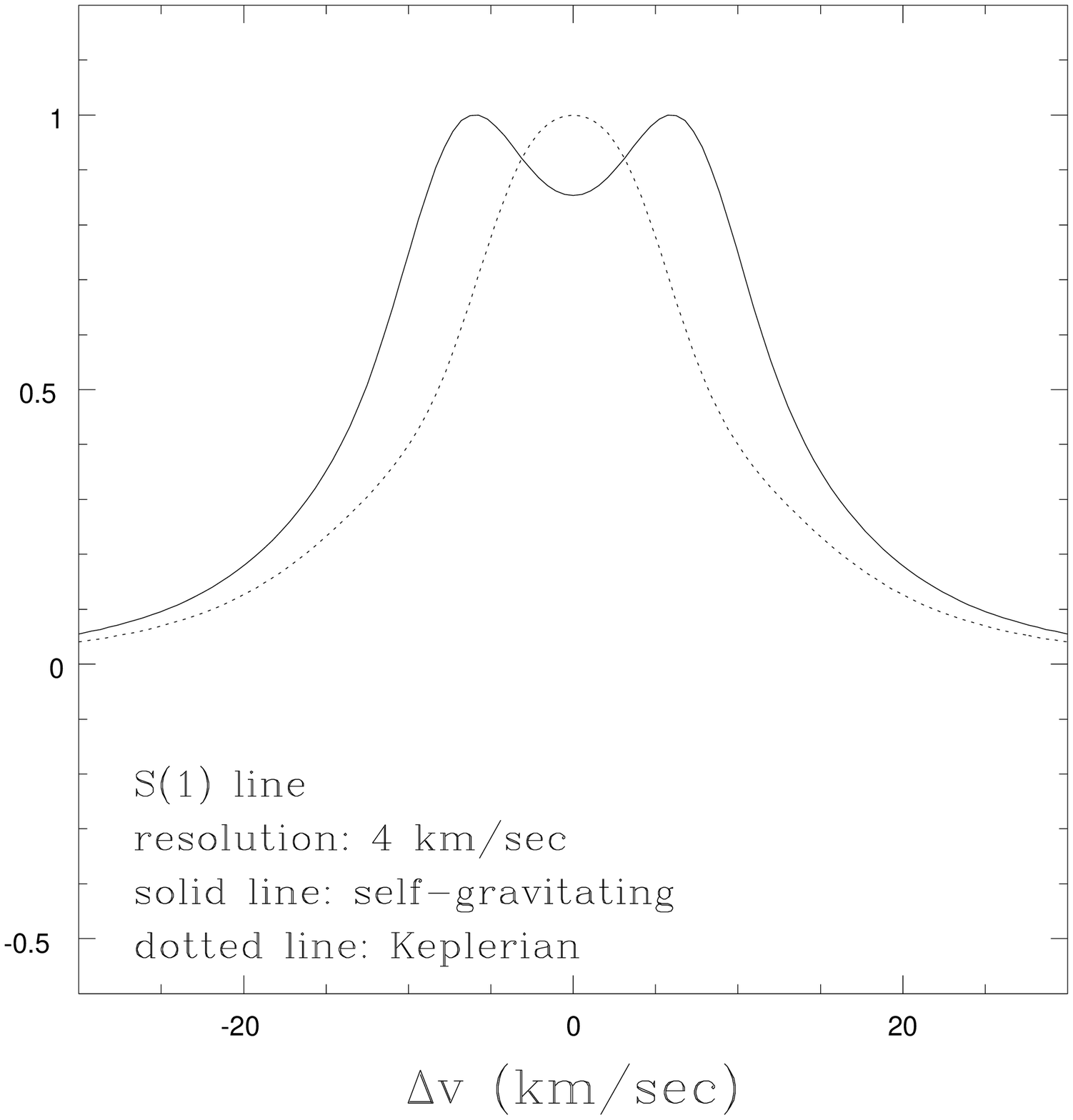,height=7.5cm}}
\caption{Molecular hydrogen absorption line profiles for disk models
with self-gravitating (with $r_s\approx 18$ AU, solid line) and
Keplerian (dotted line) rotation curve. The upper panel shows the S(0)
line at $28\micron$, with a resolution of 5km/sec ({\bf left}) and
4km/sec ({\bf right}); the lower panel shows the S(1) line at
$17\micron$, with a resolution of 5km/sec ({\bf left}) and 4km/sec
({\bf right}).}
\label{fig:mirthick}
\end{figure*}

\begin{figure*}[ht!]
\centerline{\epsfig{figure=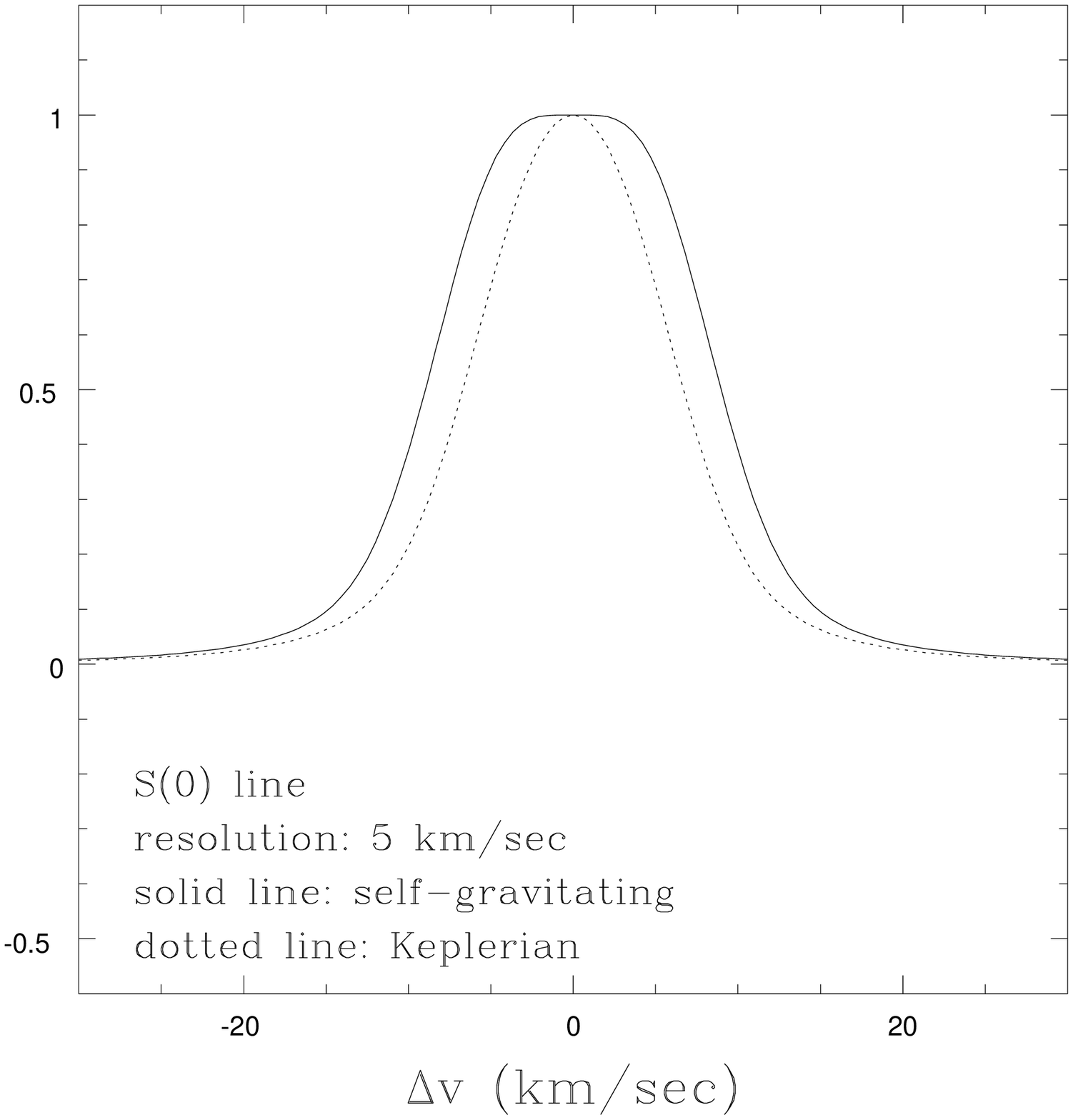,height=7.5cm}
            \epsfig{figure=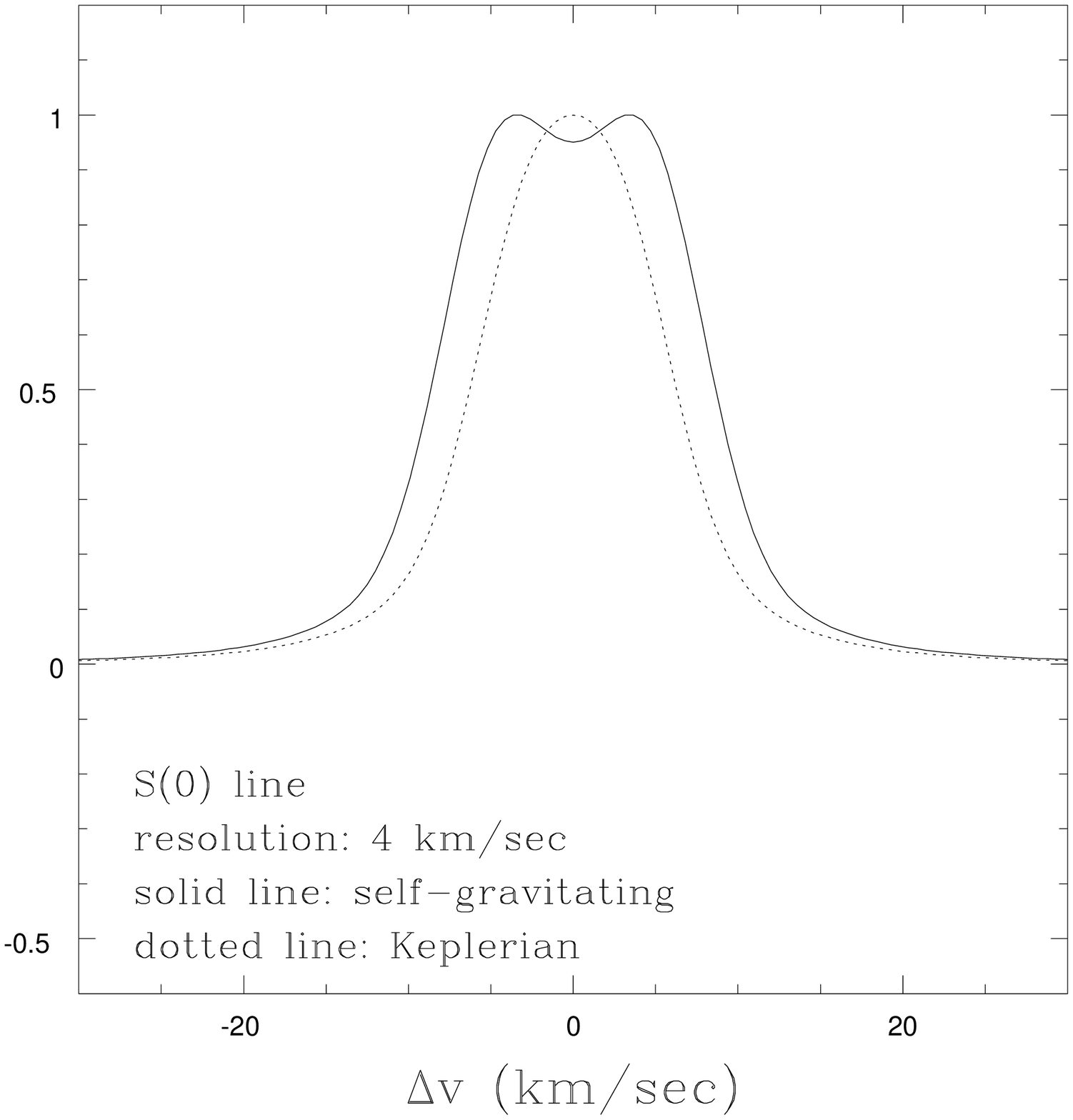,height=7.5cm}}
\centerline{\epsfig{figure=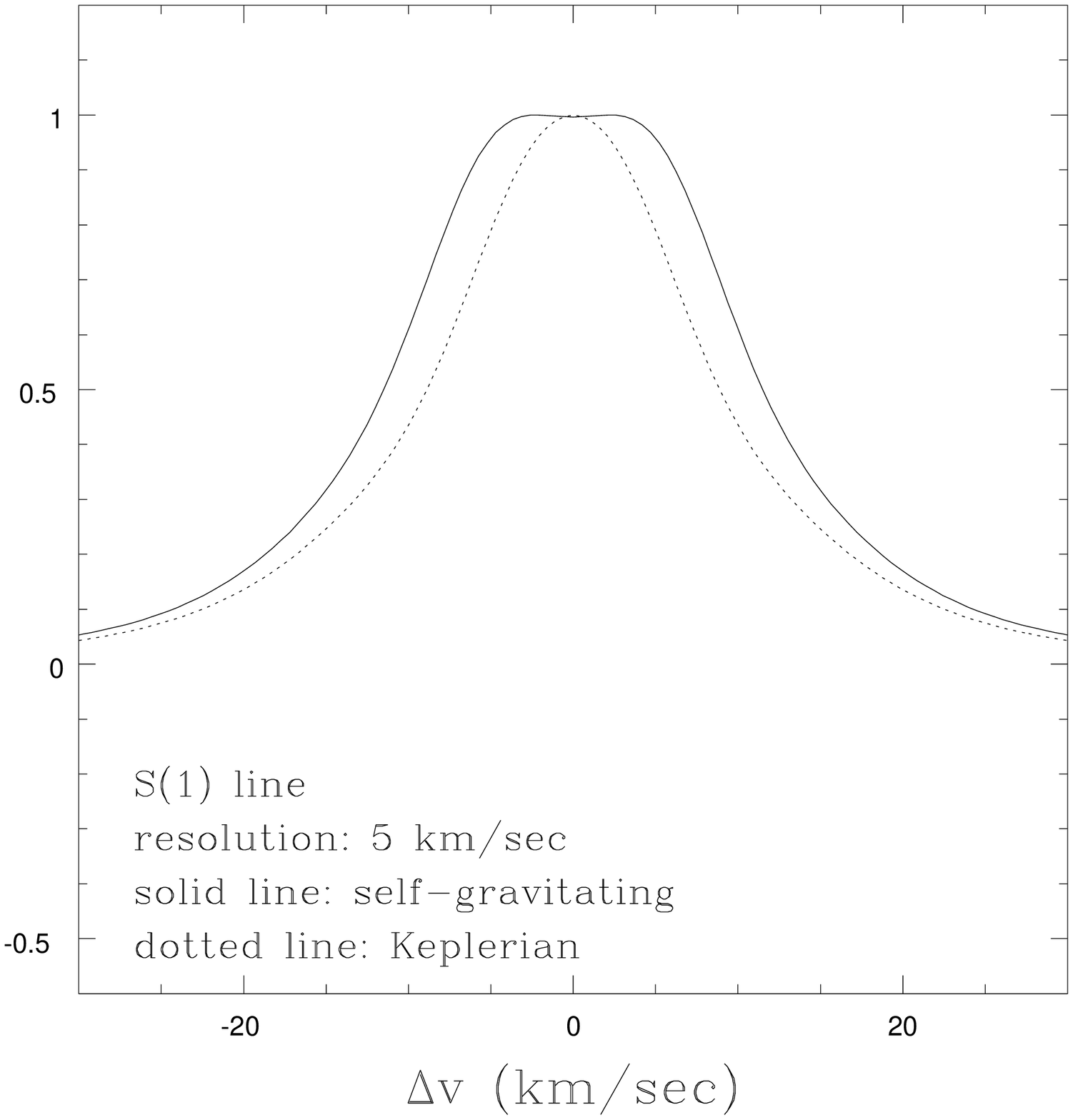,height=7.5cm}
            \epsfig{figure=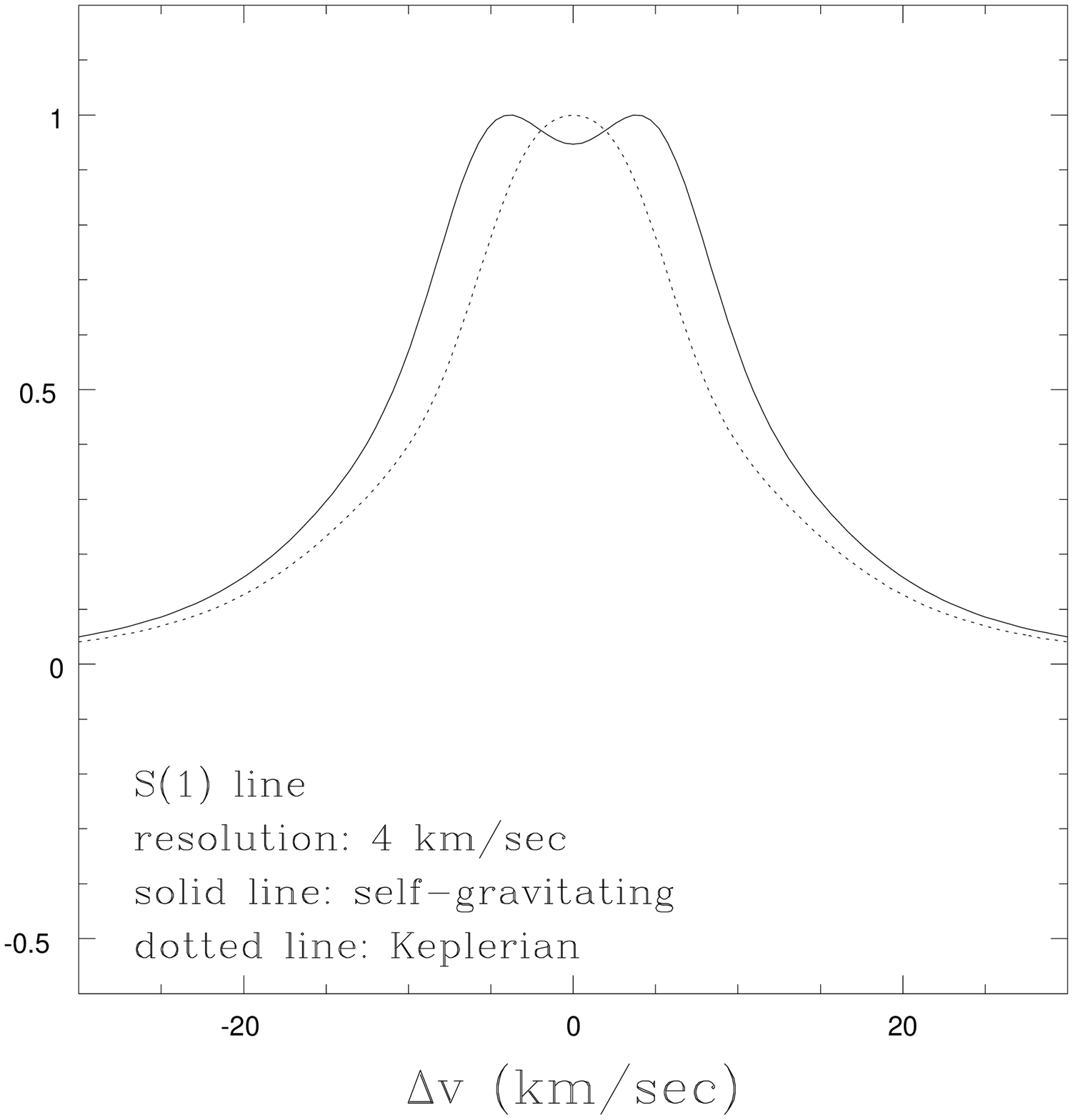,height=7.5cm}}
\caption[H$_2$ absorption line profiles, with a different $r_s$]
{Same as Fig. \ref{fig:mirthick}, but with $r_s\approx 36$ AU.}
  \label{fig:mir2thick}
\end{figure*}

A complete analysis would require the discussion of the full disk
model, with an assumed vertical temperature profile, and the solution
of the radiative transfer equation in the vertical direction, taking
into account both viscous dissipation and external irradiation as
heating sources \citep{calvet93}. Here, for simplicity, we will only
consider the two extreme cases of absorption lines (appropriate for
the case in which viscous dissipation dominates) and of emission lines
(that may be produced as an effect of irradiation of the outer disk by
the inner disk), under the hypothesis that the line emitting gas is at
the same temperature as the continuum emitting dust, i.e. that the gas
temperature is given by $T_s$ as derived from the modeling of the
SED. The line profile will be obtained following the procedure
described in Section \ref{sec:line}.

\subsection{Absorption lines}

Figure \ref{fig:mirthick} shows, for $r_s \approx 18$ AU, the
predicted line profiles for the Keplerian and non-Keplerian rotation
curves, for the S(0) and S(1) absorption lines (here plotted
``upside-down'' to allow an immediate comparison with the optical and
NIR peaks), at different spectral resolutions.  We have used the same
input parameters as for the optical and near-infrared line
profiles. The shape of the absorption line profiles is not strongly
affected by the choice of the threshold temperature above which the
equivalent width is taken to vanish. Lowering the threshold from
10000K to 1000K only leads to a slight effect on the wings of the
lines. This is because the inner regions of the disk give only a small
contribution to the line profile, due to the small area they cover.

In Fig. \ref{fig:mir2thick} we show the same line profiles, but for
the larger value of $r_s$, i.e. $r_s\approx 36$ AU. From these two
figures, it is clear that, in order to detect the small changes in the
rotation curve associated with the effects of the disk self-gravity,
we should reach very high spectral resolution (see Section
\ref{feasibility} for a discussion of the feasibility of such
observations). Another interesting point is that the S(0) and S(1)
line profiles, at the resolution of 5 and 4 km/sec, can be single
peaked in the strictly Keplerian case and double peaked in the
self-gravitating case. This would clearly distinguish between the two
different pictures, if an adequate signal-to-noise ratio could be
achieved. As expected, when a larger value of $r_s$ is assumed, the
effects of the disk self-gravity are less prominent.

\begin{figure*}[t!]
\centerline{\epsfig{figure=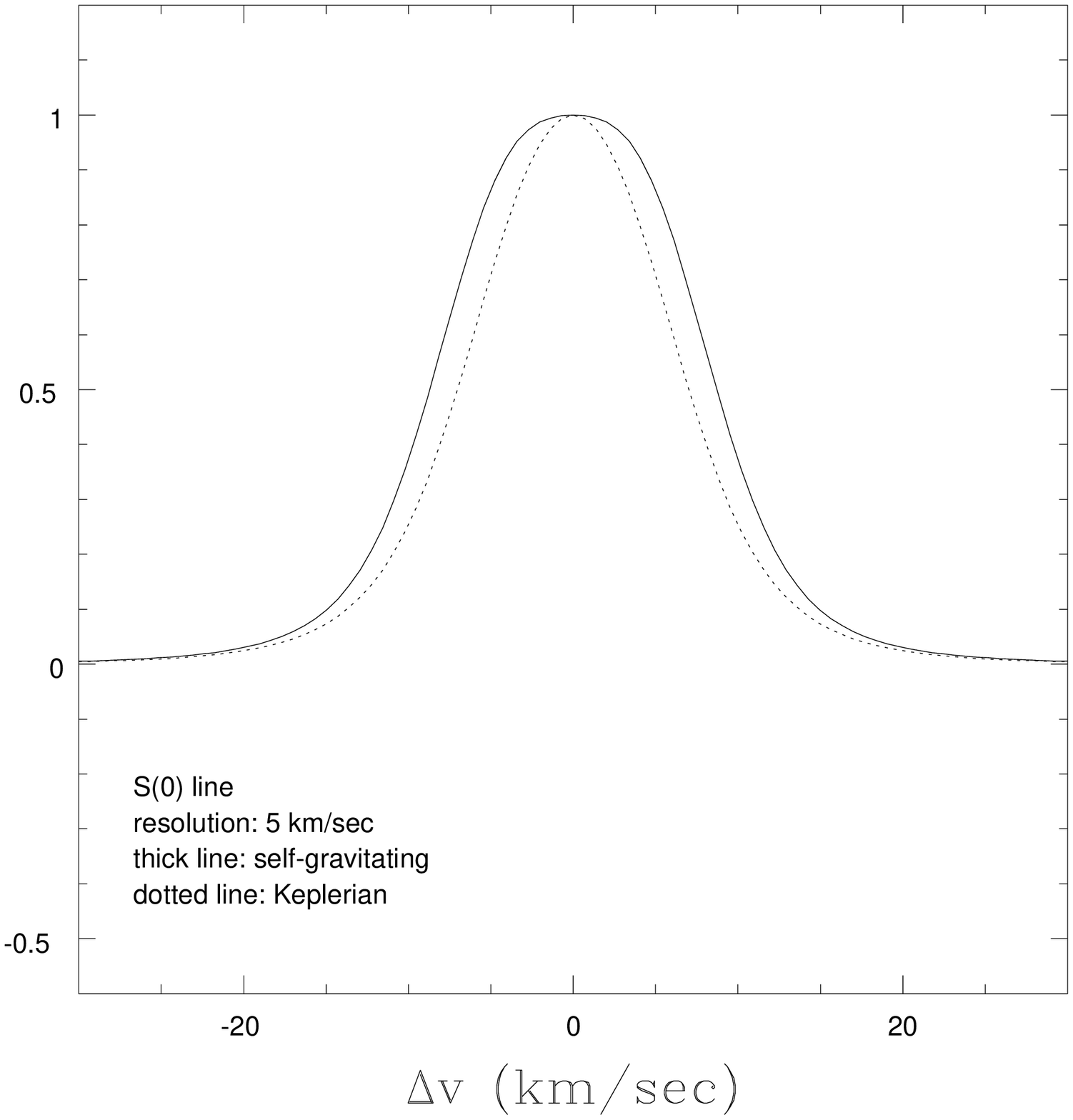,height=7.5cm}
            \epsfig{figure=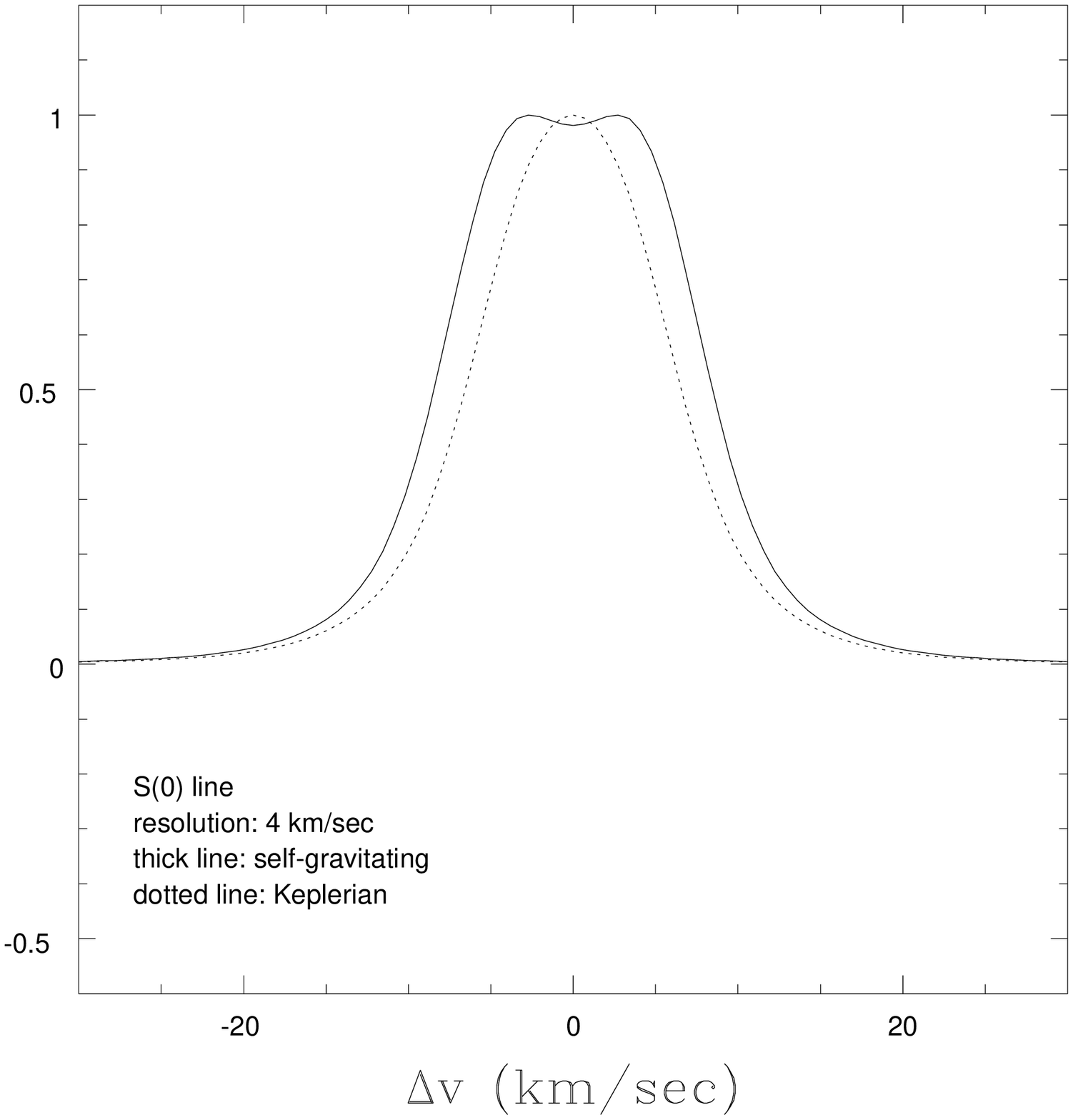,height=7.5cm}}
\centerline{\epsfig{figure=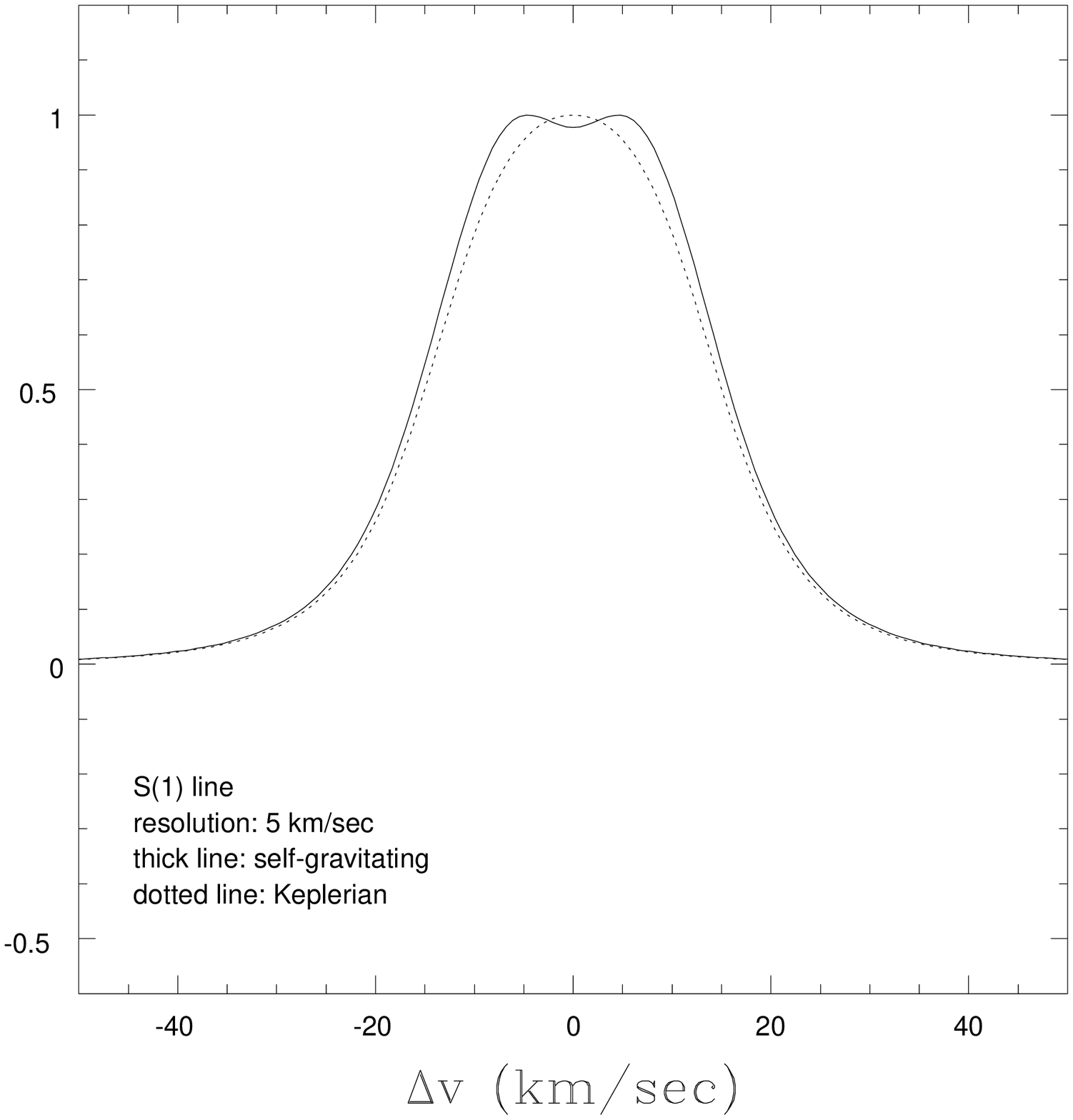,height=7.5cm}
            \epsfig{figure=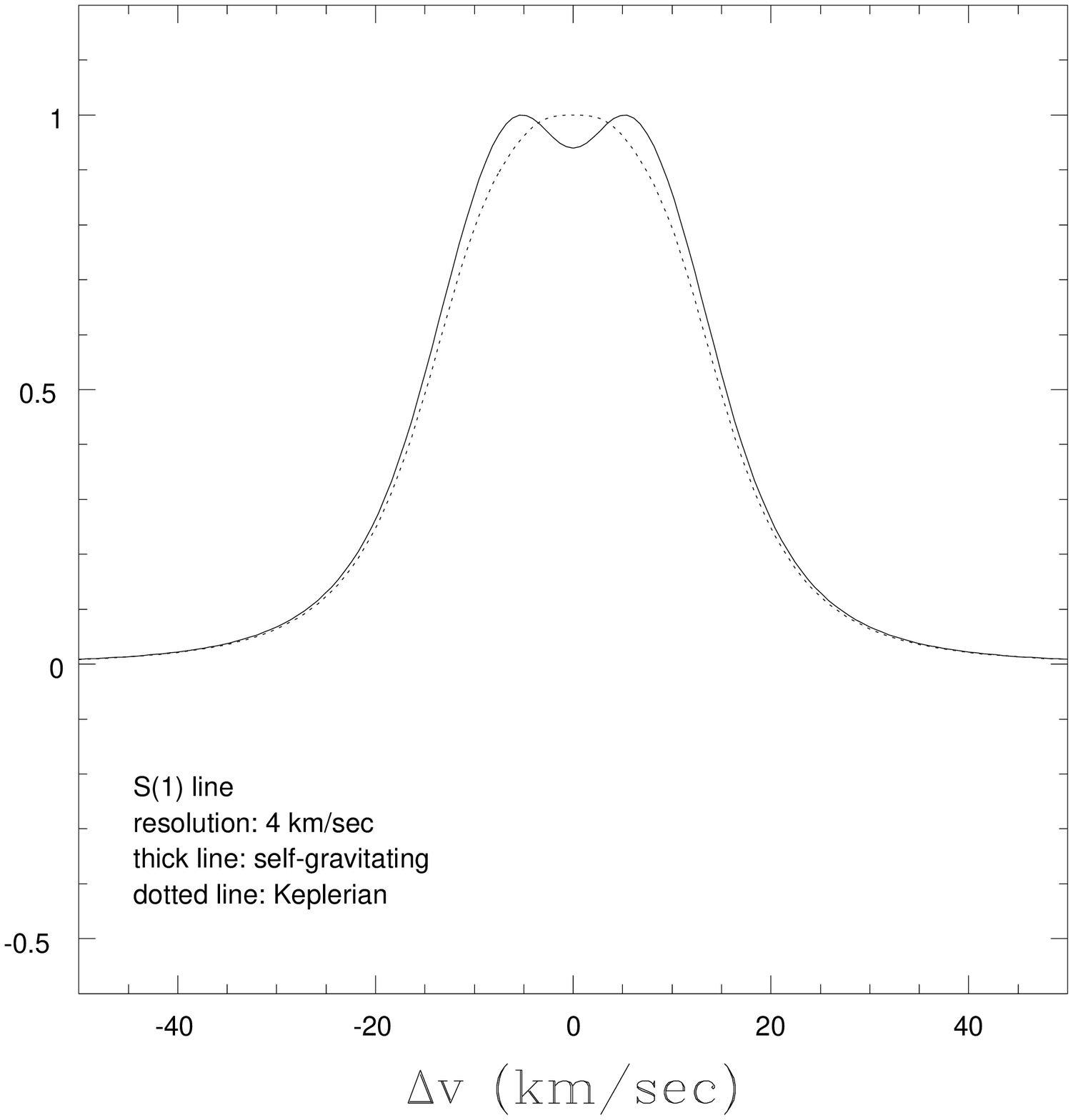,height=7.5cm}}
\centerline{\epsfig{figure=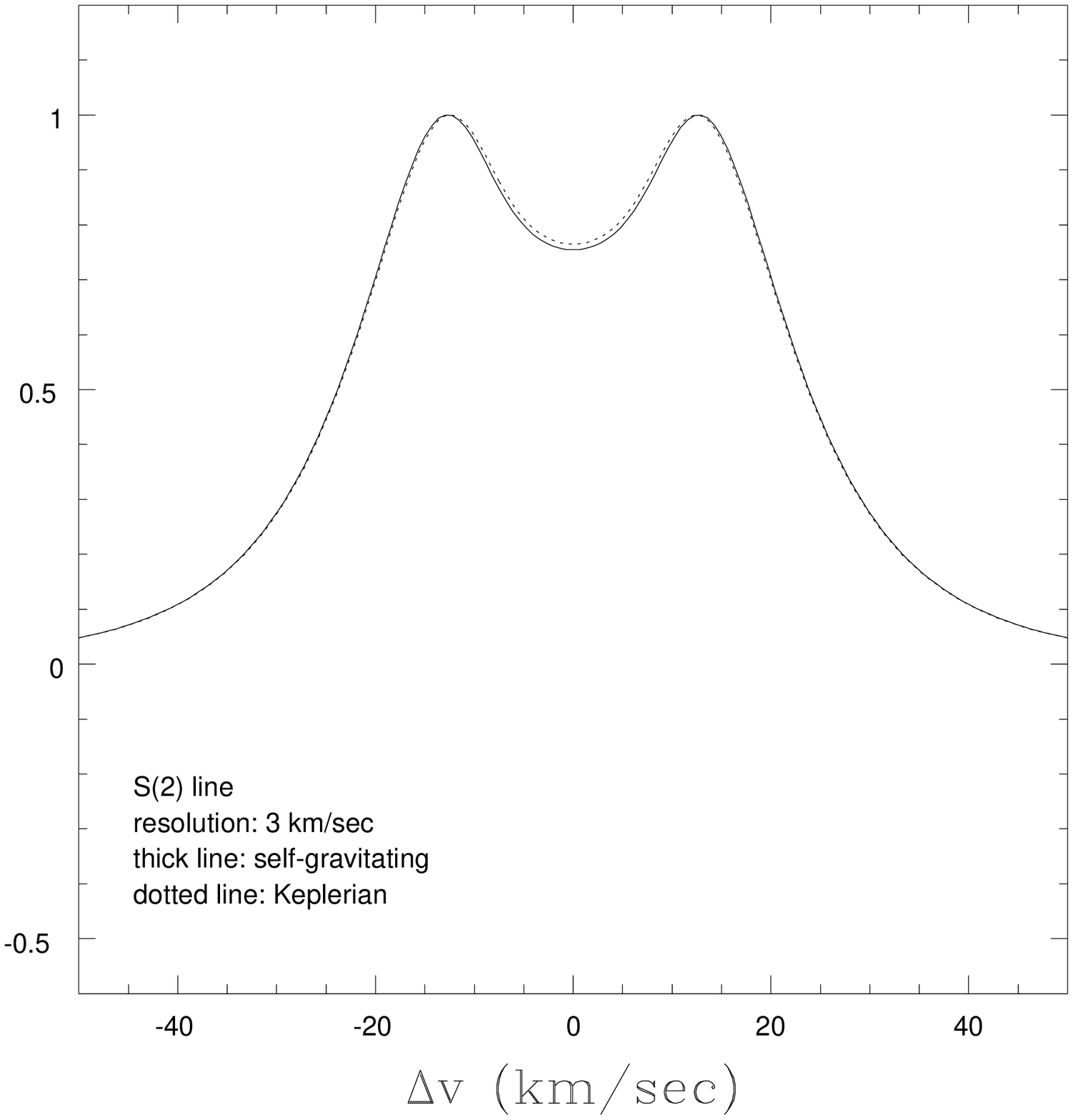,height=7.5cm}
            \epsfig{figure=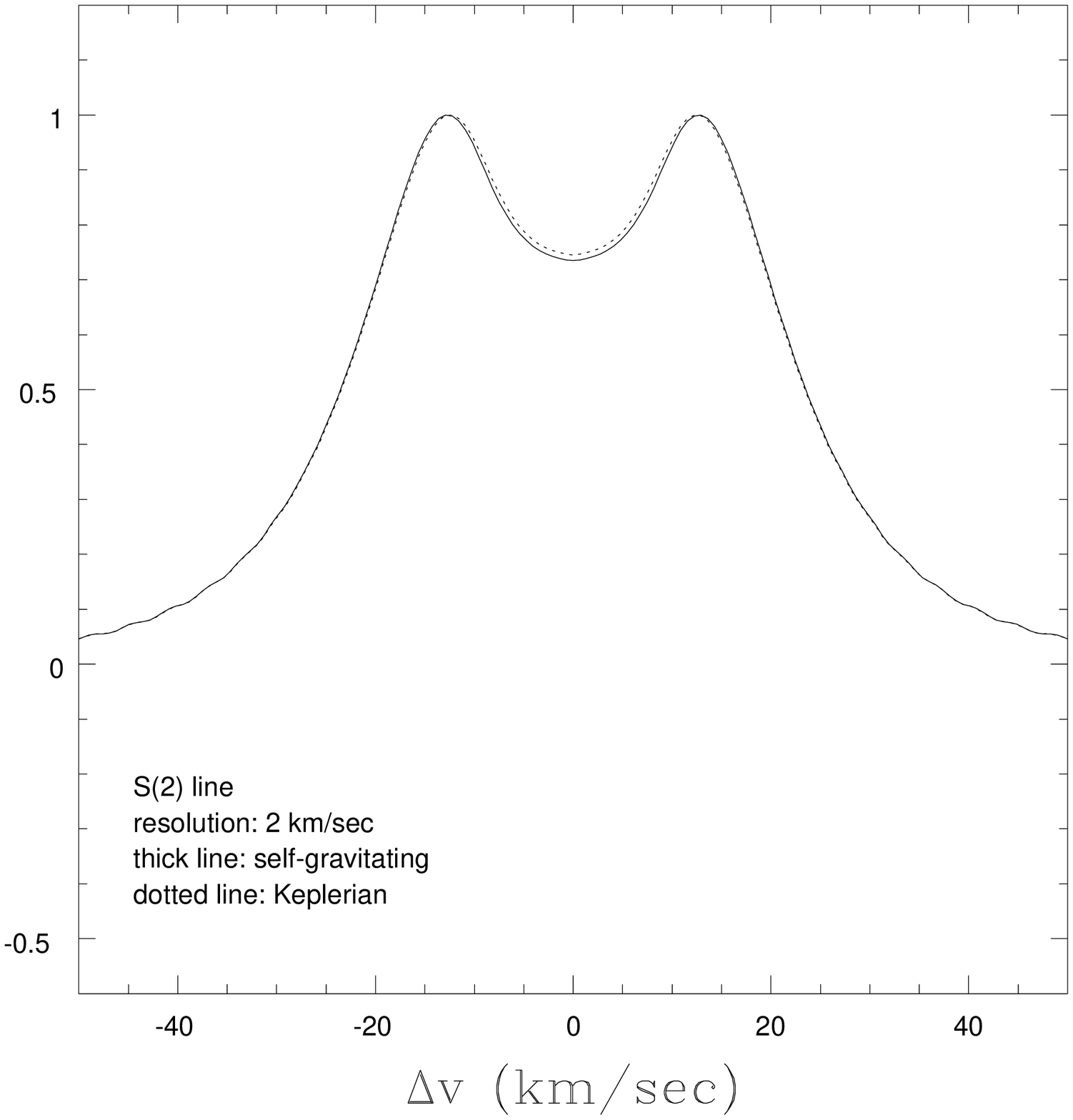,height=7.5cm}}
\caption{Molecular hydrogen emission line profiles in the
self-gravitating (solid line) case and in the strictly Keplerian
(dotted line) case, for the conservative model with $r_s \approx 36$
AU.  The upper panel shows the S(0) line at $28\micron$, with a
resolution of 5km/sec ({\bf left}) and 4km/sec ({\bf right}); the
middle panel shows the S(1) line at $17\micron$, with a resolution of
5km/sec ({\bf left}) and 4km/sec ({\bf right}); the lower panel shows
the S(2) line at $12\micron$, with a resolution of 3km/sec ({\bf
left}) and 2km/sec ({\bf right}).} \label{fig:mir}
\end{figure*}

\subsection{Emission lines}

Figure \ref{fig:mir} shows the predicted emission line profiles for
the Keplerian and self-gravitating rotation curves, for the S(0), S(1)
and S(2) lines at different spectral resolutions for the conservative
case in which $r_s\approx 36$ AU.

The general features of the shape of the lines are similar to the case
of absorption: the lines are broader in the self-gravitating case than
the corresponding lines in Keplerian disks, and they are often
double-peaked, also in cases where the Keplerian lines are
single-peaked. In the model where the transition radius $r_s$ is
smaller (not shown here), the effects of self-gravity are more
prominent, as expected.

The shape of the S(2) line profile is practically not influenced by
the modification of the rotation curve (i.e. it is mostly produced in
a region where Keplerian rotation dominates). Based on the analysis of
this and of the previous Section, we conclude that the best lines to
probe deviations from Keplerian rotation would be S(0) and S(1),
because they would display some difference between the two competing
models, with peak separation large enough to be detected with
reasonable spectral resolution.

\subsection{Feasibility}
\label{feasibility}

The above results obtained here show how deviations from Keplerian
rotation can be detected with the analysis of the shape of the
molecular hydrogen line profiles in the mid-infrared, provided that we
can achieve high spectral resolution and a high signal-to-noise
ratio. The Texas Echelon Cross Echelle Spectrograph (TEXES) operates
in the wavelength range from 5.5 to 28.5 $\micron$ and can reach a
resolution as high as 2-3 km/sec at the shortest wavelengths and of
$\approx 5$ km/sec at $17\micron$.  It has been mounted on the NASA's
3m Infrared Telescope Facility, to confirm the existence of molecular
hydrogen rotational lines apparently observed with ISO in low-mass,
non-outbursting objects, such as some T Tauri and Herbig Ae stars
\citep{richter2002b}. In addition, EXES, the companion instrument of
TEXES, is expected to be placed on SOFIA (Stratospheric Observatory
for Infrared Astronomy). Here we have demonstrated that, in principle,
these two instruments have sufficient spectral resolution to measure
this effect in FU Orionis disks.

If we consider instead the capabilities of ISO, the maximum spectral
resolution achievable is 10 km/sec.  The predicted S(0) line profiles
in FU Ori with the lower resolution of 10 km/sec is no longer
double-peaked.  In the self-gravitating case there is only a slight
broadening of the line profile with respect to the Keplerian case. We
thus consider it unlikely that the signature of non-Keplerian rotation
can be found in ISO observations.

Actually, the most serious problem with these mid-infrared line
observations is to achieve a sufficiently high signal-to-noise
ratio. For the case of emission lines, if the line emitting gas
and the continuum emitting matter are approximately at the same
temperature, we expect only weak emission lines to be present (as
confirmed by \citealt{richter2002b}). For the case of absorption,
the small absorption coefficient of the pure rotational H$_2$
transitions also leads to weak absorption lines.

\subsection{Robustness of the H$_2$ emission lines results}
\label{robustness}

The optical thickness of the relevant lines depends on the surface
density profile, which cannot be directly inferred from the modeling
of the SED alone.  Observations of protostellar disks, consistent with
our self-gravitating disk model \citepalias{BL99}, suggest a disk
density profile of the form $\sigma\propto r^{-\gamma}$, with
$\gamma\approx1$; in the previous Sections, we have indeed set
the surface density power law index to be $\gamma=1$.

\begin{figure*}[hbt!]
\centerline{\epsfig{figure=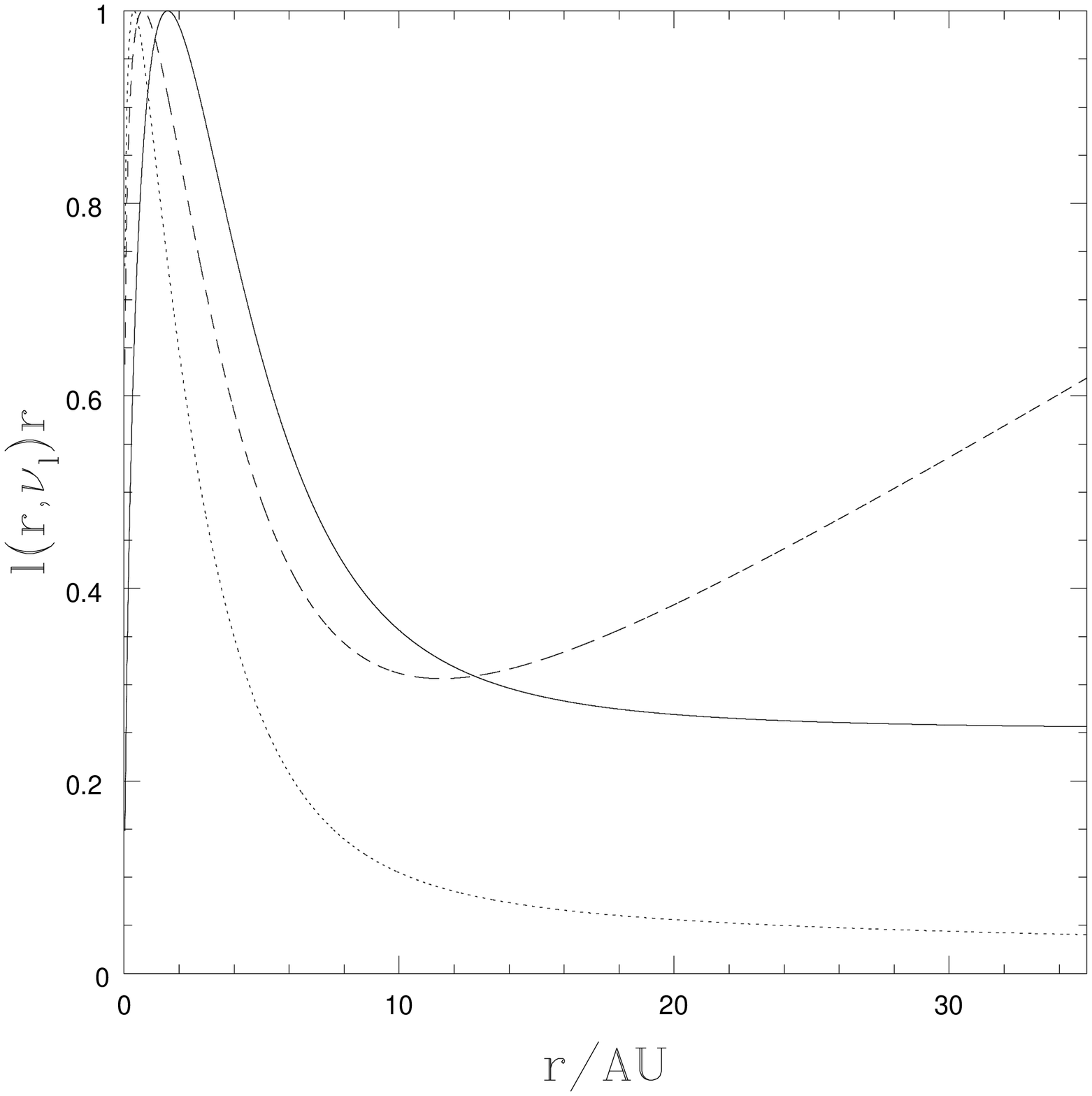,height=7cm}
            \epsfig{figure=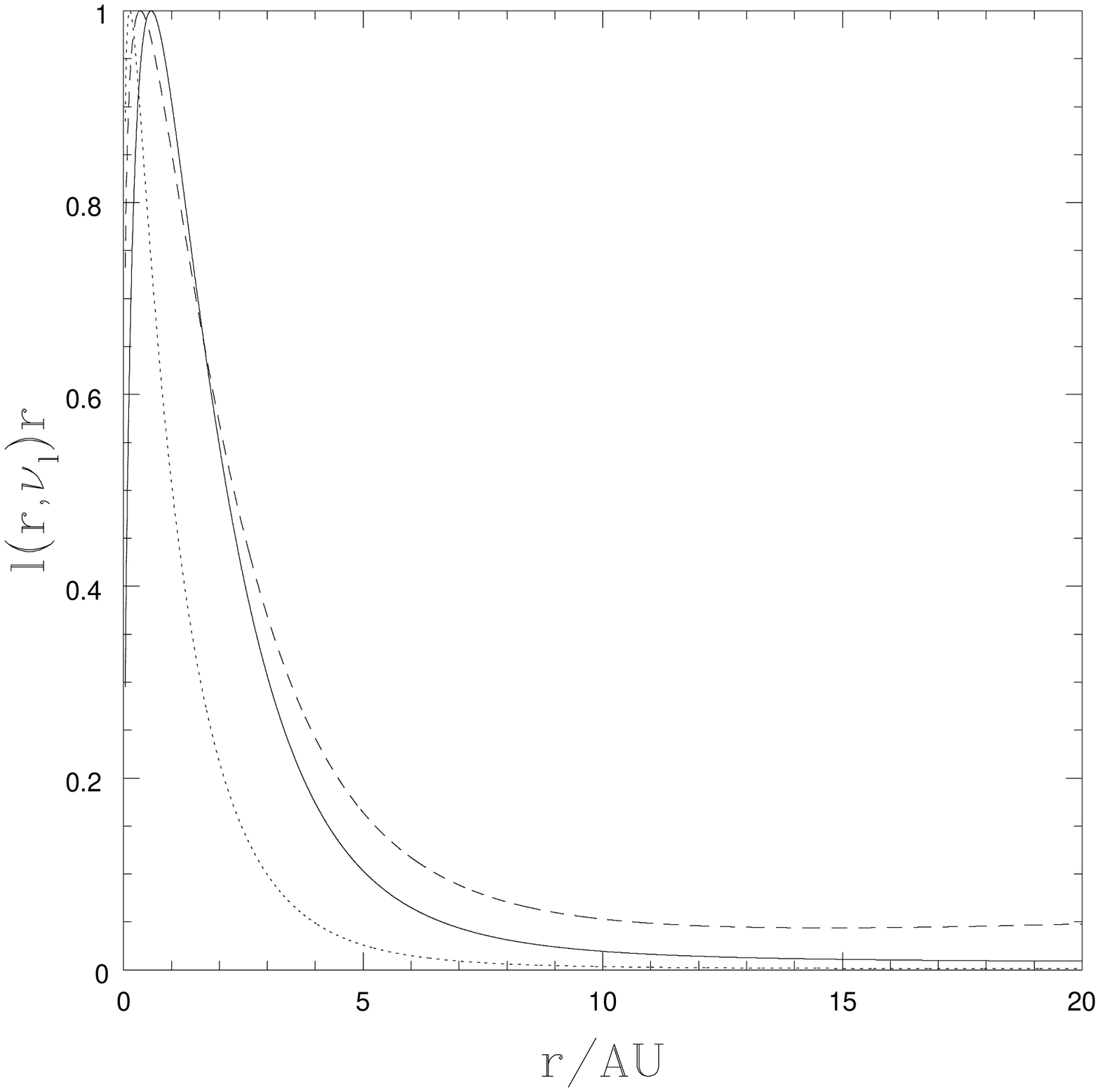,height=7cm}} 
\caption{Local emissivity as a function of radius for the S(0) (left)
and S(1) (right) line profiles. {\bf Thick solid line}: optically
thin, $\gamma=1$; {\bf Thin line}: optically thin, $\gamma=1.5$; {\bf
Dashed line}: optically thick. The small picture in the right panel is
a blow up of the profiles at small radii.} \label{fig:opa3}
\end{figure*}

The H$_2$ rotational emission line are optically thin up to relatively
high densities, due to their low Einstein's A-coefficient. FU Orionis
objects are likely to be younger, and denser, than T Tauri and
protoplanetary disks, so that for them we may even have to consider
the possibility that the H$_2$ emission is optically thick. In this
Section we discuss the robustness of the results presented in the
previous Section, with respect to the uncertainties on the optical
thickness of the disk and on its surface density profile.

In Fig. \ref{fig:opa3}, the local emissivity of every annulus of the
disk, for the S(0) and S(1) lines, is plotted as a function of radius
for three different cases, i.e.: \uno optically thin emission,
$\gamma=1$, \due optically thin emission, $\gamma=1.5$, and \tre
optically thick emission. When the steepness of the surface density
profile is increased, the peak of the emission for a certain line
occurs at relatively smaller radii, as a result of the increased
surface density. On the other hand, optically thick emission tends to
contribute more than optically thin emission at larger radii, because
in our models the temperature profile in the outer disk is
approximately constant, while the optically thin emissivity rapidly
decreases, being proportional to the surface density profile.

\begin{figure*}[hbt!]
\centerline{\epsfig{figure=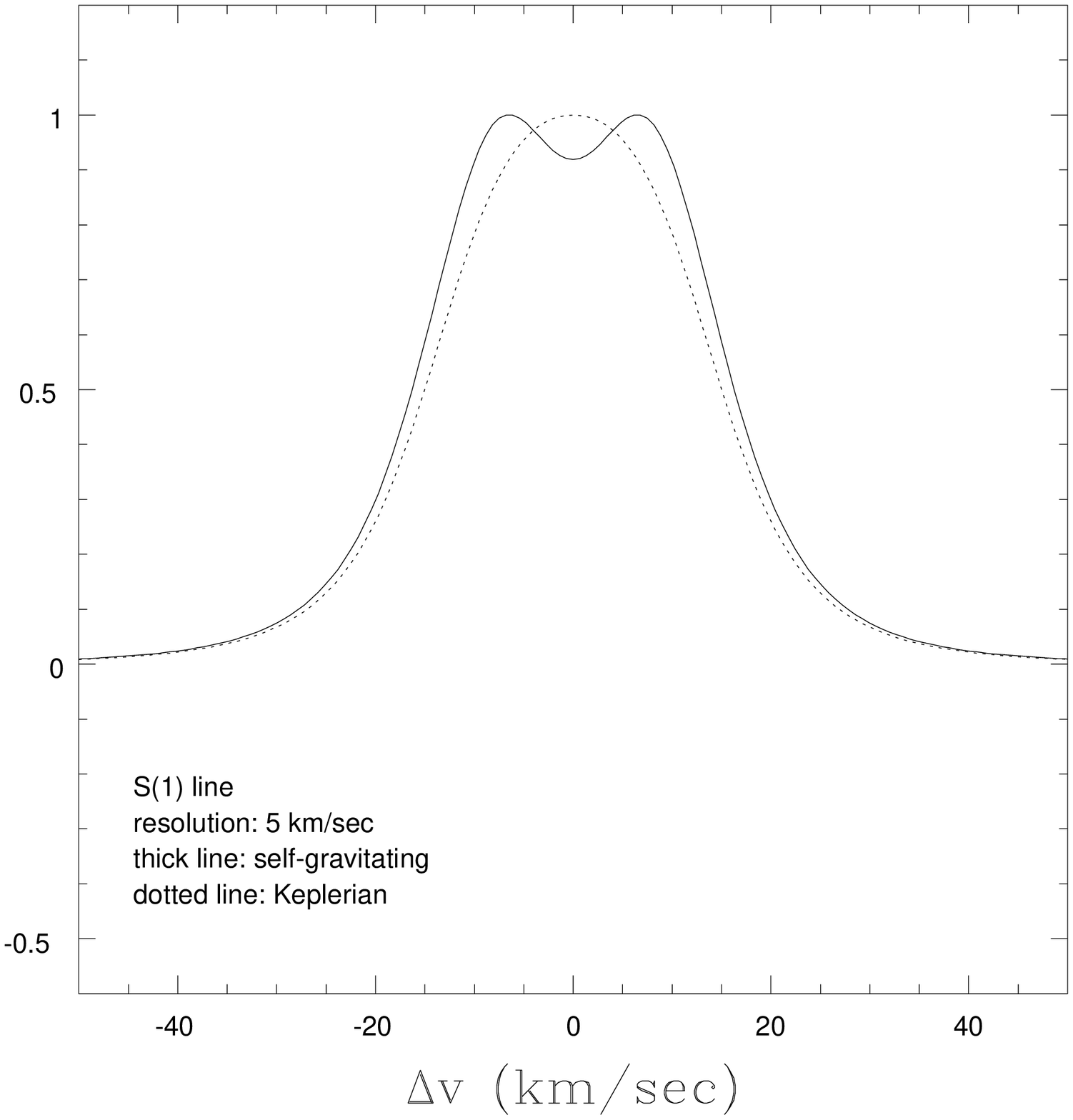,height=8cm}
            \epsfig{figure=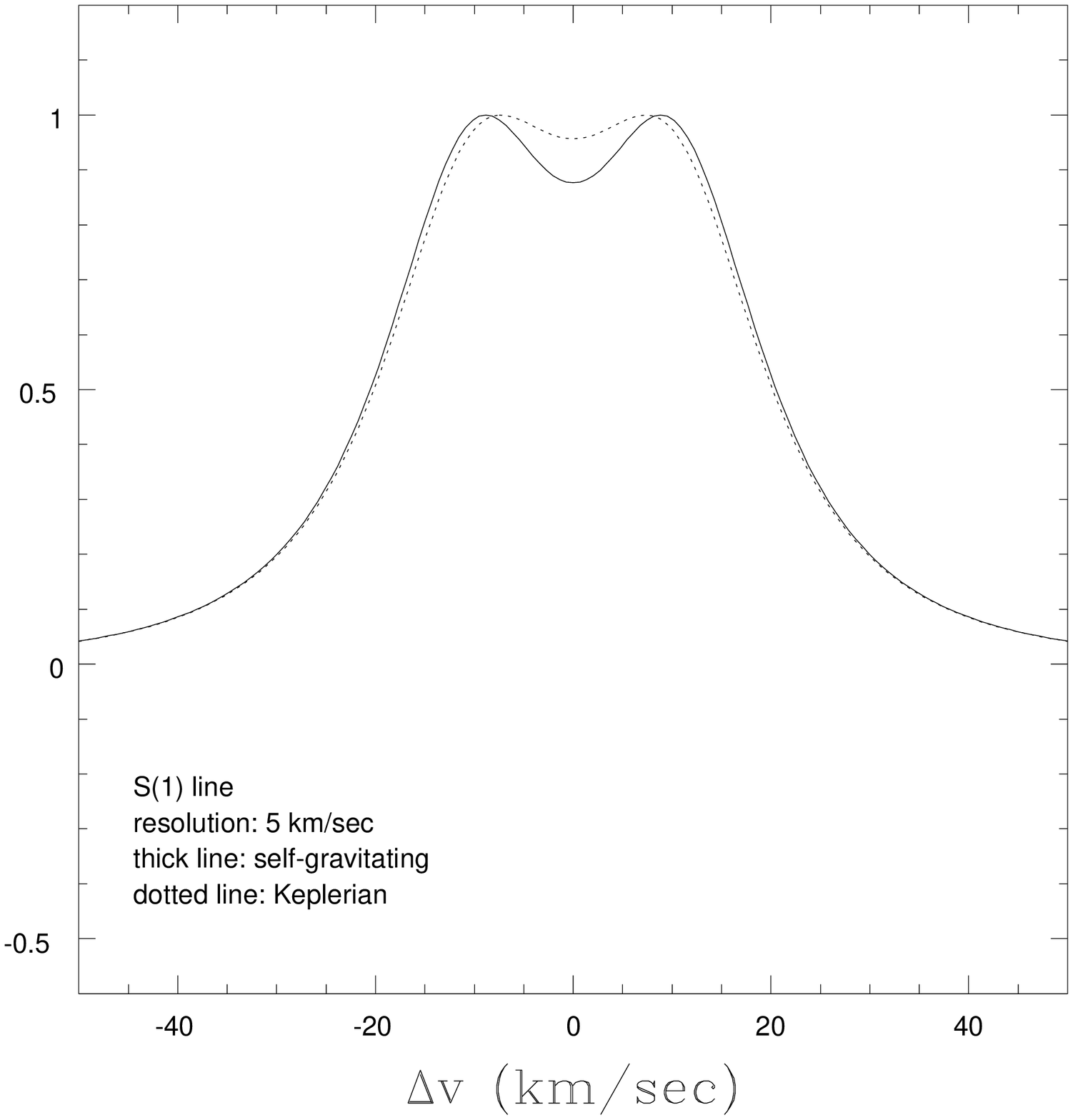,height=8cm}}
\caption{S(1) line profile in the optically thin case, with $\gamma=1$
(left) and $\gamma=1.5$ (right).}  \label{fig:opa1} 
\end{figure*}

In Fig. \ref{fig:opa1} we show how the S(1) line profile changes when
a different slope of the disk density is assumed.  A steeper density
profile (here we consider $\gamma=1.5$) results in a larger peak
separation. This is due to the fact that a steeper surface density
profile lets the inner parts of the disk, which rotate faster,
contribute more prominently, with respect to the case of a less steep
profile. This identifies a potential problem in the observational test
that we are proposing here (for the case of optically thin line
profiles). In fact, the increased contribution of the inner disk to
the line emission when a steeper surface density profile is assumed
may be such that the S(1) line is produced mostly at radii where the
rotation curve is Keplerian.  The resulting modifications to the line
profiles due to the disk self-gravity would therefore become only
barely visible in this case.

In practice, there is a sort of degeneracy in the appearance of
optically thin line profiles with respect to the assumed model: a
broader line profile can be traced either to the effects of the disk
self-gravity, which enhances the rotation at large radii, or to the
steepness of the disk density profile, which makes it possible for a
certain emission line to originate from smaller radii, where the
rotation velocity is larger.

\section{Sub-mm CO line profiles}
\label{radio}

The discussion of the previous Section shows that some H$_2$ emission
lines can be a useful tool to probe the rotation of the outer disk,
although confusion may arise in the interpretation of optically thin
lines because, under certain circumstances, deviations from the
typical $\sigma \sim 1/r$ disk density profile can mimick those
associated with deviations from Keplerian rotation.

\begin{figure*}[hbt!]
\centerline{\epsfig{figure=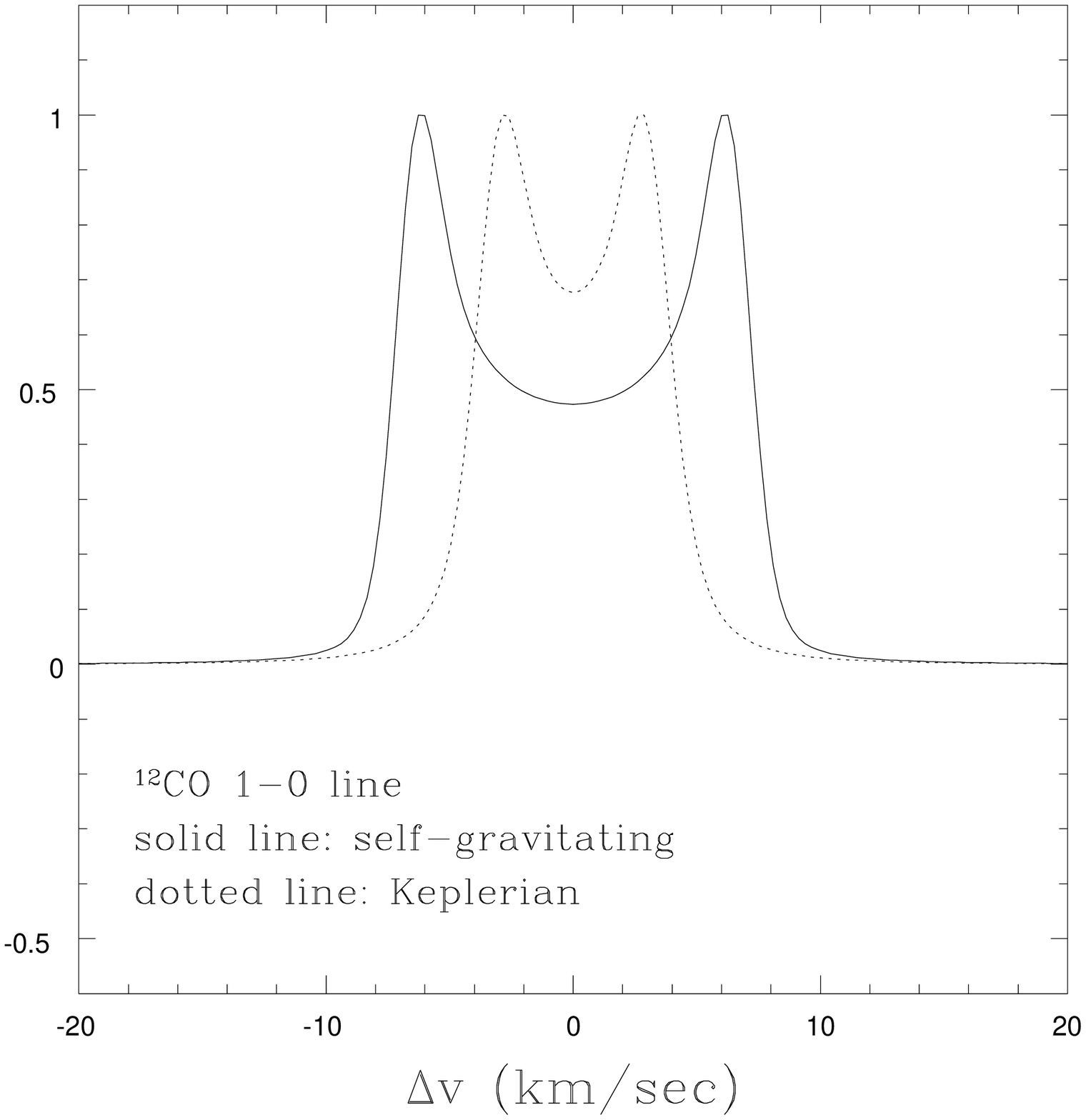,height=8cm}
            \epsfig{figure=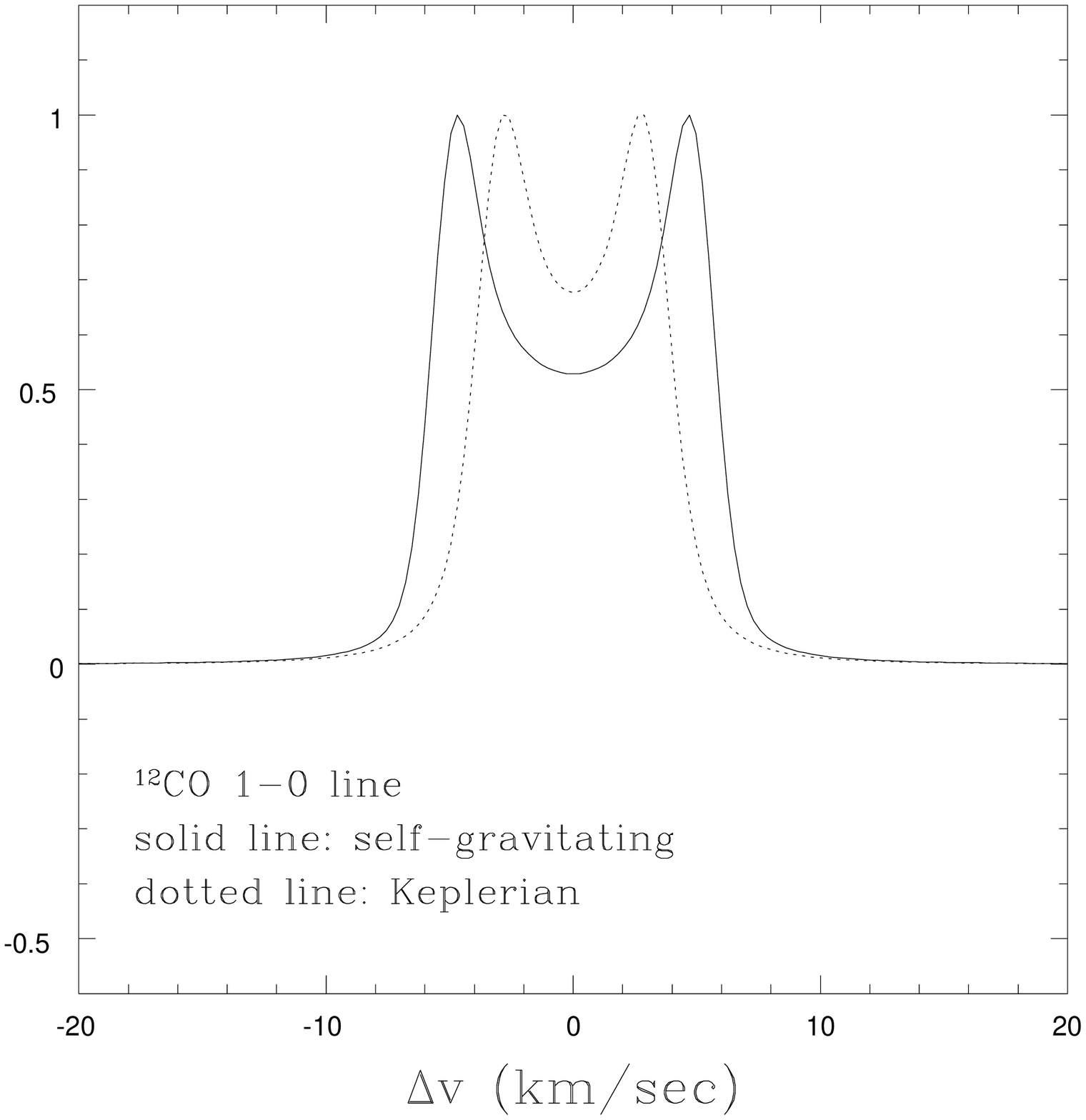,height=8cm}} 
\caption{$^{12}$CO 1-0 line profile at 110 GHz. {\bf Solid line}: for
a model applicable to FU Ori with a non-Keplerian rotation curve
associated with a self-gravitating accretion disk; {\bf Dotted line}:
for a strictly Keplerian model of FU Ori. The results are shown for
$r_s\approx 18$ AU (left) and $r_s\approx 36$ AU (right).}
\label{fig:co}
\end{figure*}

To derive more convincing conclusions about the rotation of the outer
disk, it would thus be important to consider long-wavelength lines
that are optically thick, so that the additional assumptions
concerning the surface density profile are not needed. In this
respect, some sub-mm CO lines are very interesting, because they are
expected to be optically thick \citep{becksarg93,thi2001} and can be
studied with very high spectral resolution. These global CO line
profiles, observed in T Tauri and protoplanetary disks by a number of
authors, usually show the typical double-peaked profile
\citep{thi2001,dutrey97}.

Interestingly, millimetric CO line emission has also been observed in
V1331 Cyg \citep{mcmuldroch93}, an ``FU Orionis object between
outbursts'', which shares many features with FU Orionis objects, but
has a smaller luminosity. The estimated disk mass in V1331 Cyg is
$\approx 0.5\msun$, a further clue to the fact that FU Orionis disk
mass may be fairly high.

As an example, we consider the $^{12}$CO 1-0 at 110 GHz, which is
commonly observed in protostellar disks. We assume that the main
limiting factor for the spectral resolution is given by turbulent or
thermal broadening and we take the factor $v_b$ in the Gaussian with
which the line profile is convolved (see Section \ref{sec:line}) to be
1 km/sec. The resulting line profiles, for the two different choices
of $r_s$, are shown in Fig. \ref{fig:co}. The effect of self-gravity
is clear in both cases.

In any case, as noted in the Introduction, care should be taken
(especially for younger objects, such as FU Ori) in order to properly
identify the various kinematical components (such as winds and
outflows) that might contribute to the shape of the line
profile. Another source of uncertainty related to CO lines is that of
molecular depletion, resulting from the freezing out of molecular
species onto dust grains. Many studies (see, for example,
\citealt{markwick02}) have shown that, in the outer and colder disk,
the column density of many species can be reduced significantly.
However, these studies refer mostly to parameters appropriate for T
Tauri disks, while we expect that FU Orionis disks, which are
characterized by a much higher temperature, will be less subject to
depletion.

\section{Conclusions}
\label{conclusion}

The study of the line profiles produced in Young Stellar Objects has
traditionally played a very important role in determining the physical
conditions of circumstellar matter. In particular, it has been used to
confirm the accretion disk origin of the luminosity of FU Orionis
objects \citepalias{kenyon88} and to test the Keplerian rotation in
many T Tauri stars \citep{guilloteau94}. In addition, spectroscopic
observations may be able to probe the vertical structure of the disk
(see, for example, \citealt{cheng88} for the disks in dwarf novae).

{\it Spatially resolved} sub-mm interferometric observations of T
Tauri stars have already provided interesting information about the
outer disk kinematics \citep{guilloteau98,simon01}. Such radio
observations can reach an angular resolution of $\approx
1\arcsec$. However, for objects such as FU Ori, which lies at a
distance of $\approx 500$ pc, the angular size of the disk (which
should extend out to 50-60 AU, judging from the modeling of the SED)
is likely to be as small as $\approx 0.1\arcsec$.

In this paper we have exploited the wavelength dependence of the
radial position of the peak of the line emissivity of the disk, to
obtain additional information on the kinematical properties of the
outer disk in FU Orionis objects, based on the shape of {\it global
line profiles}. Our main purpose has been to define a clear cut test
about the importance of the disk self-gravity in determining the
rotation curve of the accretion disk in the early stages of star
formation.

The method presented in this paper is based on the complementary
information that can be obtained from the analysis of the optical and
near-infrared line profiles (which probe the inner rotation curve, and
hence set the value of the mass of the central object) and the
analysis of high spectral resolution line profiles at long-wavelengths
(which probe the rotation curve in the outer disk). This is a
generalization of the method discussed by \citet{kenyon88} and
\citet{popham96}.

The high spectral resolution that can be obtained with the ground
based and already operating TEXES instrument, and with its air-borne
companion instrument EXES that will be on board of the SOFIA
observatory \citep{richter2002b}, may indeed give us the way to
distinguish between Keplerian and non-Keplerian rotation based on the
analysis of pure rotational H$_2$ lines. This in an important test
about the viability of the self-gravitating disk model in FU Orionis
objects, independent of the results of the SED modeling by
\citet{LB2001}. In addition, once we accept the self-gravitating disk
model, this test will give us an alternative, independent way to
measure the viscosity parameter $\alpha$, thus providing insights into
the physical mechanisms that operate in protostellar disks.

As far as the shape of optically thin line profiles is concerned, we
have also considered the robustness of our conclusions with respect to
changes in the surface density profile of the disk. We have found that
changes in the surface density profile could in principle mimick the
effects obtained by changing the rotation curve of the disk. In the
present study we have adopted a very simplified model for the vertical
structure of the disk. More realistic models of the radiative transfer
in the vertical direction would be needed to address this issue
further.

An additional difficulty with the use of the proposed diagnostics for
mid-infrared lines is related to the fact that an emission line at
wavelengths where the dust disk is optically thick (such as the lines
in the infrared wavelength range) would be produced only if there is a
spatial or temperature separation between the warm gas and the
dust. In this respect, the recent non-detection of H$_2$ rotational
lines from a small sample (two Herbig Ae stars and one T Tauri star)
of pre-main-sequence stars by \citet{richter2002b} suggests that gas
and dust are not spatially separated, as would result in some flaring
disk model \citep{chiang97}. This may indeed be the case for the
highly accreting FU Orionis objects that we consider here.

Finally we have come to the conclusion that optically thick,
sub-millimetric line profiles, such as those of some CO lines, often
observed in protostellar disks, are basically free from the
difficulties noted for the mid-infrared lines and thus represent the
best tool to probe the physical conditions of the outer parts of the
massive accretion disks in FU Orionis objects.

\begin{acknowledgements}
GL wishes to acknowledge the support of the Scuola Normale Superiore
in Pisa, where most of this work has been carried out. This work has
been partially supported by MIUR of Italy.
\end{acknowledgements}

\bibliographystyle{aa}
\bibliography{lodatoaa}

\end{document}